\documentclass[useAMS,usenatbib]{mn2e}
\usepackage{graphicx}
\usepackage{amssymb} 

\newcommand\about     {\hbox{$\sim$}}

\def\eq#1{\begin{equation} #1 \end{equation}}

\def\mic              {\hbox{$\mu{\rm m}$}}

\def\U                {{\em uz}}
\def\V                {{\em vz}}
\def\B                {{\em bz}}
\def\Y                {{\em yz}}

\def\z                {{z}}
\def\BY               {\hbox{$bz-yz$}}
\def\VY               {\hbox{$vz-yz$}}
\def\UV               {\hbox{$uz-vz$}}

\def\mic              {\hbox{$\mu{\rm m}$}}
\def\comm#1  {{\tt (COMMENT: #1)}}

\title[The Rest-frame Colors of SDSS Galaxies] {The Rest-frame Optical Colors of 99,000 SDSS Galaxies}

\author[V. Smol\v{c}i\'{c} et al.]{V. Smol\v{c}i\'{c}$^{1,2,3}$, \v{Z}. Ivezi\'c$^{4,1}$,
M. Ga\'{c}e\v{s}a$^{2,5}$, K. Rakos$^{6}$, K. Pavlovski$^{2}$, S. Iliji\'{c}$^{7}$,
\newauthor  
M. Obri\'{c}$^{8,4}$, R.H. Lupton$^{1}$, D. Schlegel$^{1}$, G. Kauffmann$^{9}$,
C. Tremonti$^{10}$, \newauthor %
J. Brinchmann$^{9}$, S. Charlot$^{9}$, T.M. Heckman$^{11}$, G.R. Knapp$^{1}$, J.E. Gunn$^{1}$,
\newauthor %
J. Brinkmann$^{12}$, I. Csabai$^{13}$, M. Fukugita$^{14}$,  J. Loveday$^{15}$\\  
$^{1}$Princeton University Observatory, Peyton Hall, Princeton NJ 08544-1001, USA\\
$^{2}$University of Zagreb, Physics Department, Bijeni\v{c}ka cesta 32, 10000 Zagreb, Croatia\\
$^{3}$Max Planck Institut f\"ur Astronomie, K\"onigstuhl 17, Heidelberg, D-69117, Germany\\
$^{4}$Department of Astronomy, University of Washington, Box 351580, Seattle, WA 98195-1580, USA\\
$^{5}$University of Connecticut, Physics Department, 2152 Hillside Road, Storrs, CT 06269-3046, USA\\
$^{6}$Institut f\"ur Astronomie, Universit\"at Wien, T\"urkenschanstrasse 17, A-1180 Wien, Austria\\
$^{7}$Faculty of Electrical Engineering and Computing, Unska 3, 10000 Zagreb, Croatia\\
$^{8}$Kapteyn Astronomical Institute, University of Groningen, P.O.BOX 800, 9700AV Groningen, The Netherlands\\
$^{9}$Max-Planck-Institute f\"ur Astrophysik, D-85748, Garching, Germany\\
$^{10}$Hubble Fellow, University of Arizona, Steward Observatory, 933 N. Cherry Ave., Tucson, AZ 85721, USA\\
$^{11}$Department of Physics \& Astronomy, Johns Hopkins University, Baltimore, MD 21218, USA\\
$^{12}$Apache Point Observatory, 2001 Apache Point Road, P.O. Box 59, Sunspot, NM 88349-0059, USA\\
$^{13}$Department of Physics, E\"otv\"os University, Postfach 32, Budapest H-1518, Hungary\\
$^{14}$Institute for Cosmic Ray Research, University of Tokyo, 5-1-5 Kashiwa, Kashiwa City, Chiba 277-8582, Japan\\
$^{15}$Astronomy Centre, University of Sussex, Falmer, Brighton, BN1 9QJ}

\begin{document}
\maketitle
\label{firstpage}

\begin{abstract}
We discuss the colors of 99,088 galaxies selected from the Sloan Digital Sky Survey Data Release 1 
``main'' spectroscopic sample (a flux limited sample, $r_{Pet}<17.77$, for 1360 deg$^2$) in the
rest-frame Str\"omgren system (\U, \V, \B, \Y). This narrow-band (\about 200~\AA) photometric system, 
first designed for the determination of effective temperature, metallicity and gravity of stars, 
measures the continuum spectral slope of galaxies in the rest-frame 3200--5800~\AA\ wavelength range. 
We synthesize rest-frame Str\"omgren magnitudes from SDSS spectra, and find that galaxies form a 
remarkably narrow locus (\about 0.03~mag) in the resulting color-color diagram. Using the Bruzual~\& Charlot 
population synthesis models, we demonstrate that the position of a galaxy along this locus
is controlled by metallicity and age of the dominant stellar population. The distribution of  
galaxies along the locus is bimodal, with the local minimum corresponding to an \about 1~Gyr
old single stellar population. The position of a galaxy perpendicular to the locus is independent
of metallicity and age, and reflects the galaxy's dust content, as implied by both the models and the 
statistics of IRAS detections.

Comparison of the galaxy locus in the rest-frame Str\"omgren color-color diagram with the galaxy 
locus in the $H_\delta$-$D_n(4000)$ diagram, utilized by Kauffmann et al.~(2003) to estimate stellar 
masses, reveals a tight correlation, although the two analyzed spectral ranges barely overlap. 
Furthermore, the rest-frame $r-i$ color (5500--8500~\AA\ wavelength range) can be ``predicted''
with an rms of 0.05~mag using the rest-frame Str\"omgren colors. {\it Overall, the galaxy 
spectral energy distribution in the entire UV to near-IR range can be described as a single-parameter 
family with an accuracy of 0.1~mag, or better.} This nearly one-dimensional
distribution of galaxies in the multi-dimensional space of measured parameters strongly supports 
the conclusion of Yip et al. (2004), based on a principal component analysis, that SDSS galaxy 
spectra can be described by a small number of eigenspectra. Furthermore, the
rest-frame Str\"omgren colors correlate tightly with the classification scheme
proposed by Yip et al. (2004) based on the first three eigenspectra. Apparently, the contributions of
stellar populations that dominate the optical emission from galaxies are combined in a simple 
and well-defined way. We also find a remarkably tight correlation between the rest-frame 
Str\"omgren colors of emission-line galaxies and their position in the Baldwin-Phillips-Terlevich 
diagram. These correlations between colors and various spectroscopic diagnostic parameters support 
earlier suggestions that rest-frame Str\"omgren photometry offers an efficient tool to study faint 
cluster galaxies and low surface brightness objects without performing time-consuming spectral 
observations.
\end{abstract}


\section{Introduction}

The studies of galaxies have been recently invigorated due to the advent of modern
sensitive large-area surveys across a wide wavelength range. The Sloan Digital Sky
Survey (SDSS, York et al. 2000, Stoughton et al. 2002, Abazajian et al. 2003) 
stands out among these surveys because it has already provided UV to near-IR five-color imaging 
data, and high-quality spectra for over 100,000 galaxies. The spectroscopic 
galaxy sample is defined by a simple flux limit (Strauss et al. 2002), and will eventually 
include close to 1,000,000 galaxies. 

A number of detailed galaxy studies based on SDSS data have already been published.
Strateva et al. (2001) and Shimasaku et al. (2001) demonstrated a tight correlation
between the $u-r$ color, concentration of the galaxy's light profile as measured
by the SDSS photometric pipeline {\it photo} (Lupton et al. 2002), and morphology.
Blanton et al. (2001) presented the SDSS galaxy luminosity function, and Kauffmann 
et al. (2003ab) determined and analyzed stellar masses and star-formation histories 
for 100,000 SDSS galaxies. A  large number of other, published and ongoing, studies are 
based on the rich dataset provided by SDSS.  

SDSS galaxy spectra have high quality (R$\sim$1800, 3800-9200~\AA\ wavelength range,
spectrophotometric calibration better than 10\%; for more details see Stoughton 
et al. 2002), and have been used in a number
of detailed investigations, including studies of elliptical (Eisenstein et al. 2003,
Bernardi et al. 2003abcd), star-forming (Hopkins et al. 2003), and active galaxies 
(Kauffmann et al. 2003c, Hao et al. 2005ab, Zakamska et al. 2003, 2004). Yip et al. 
(2004) analyzed the SDSS spectra
of 170,000 galaxies using principal component analysis (Karhunen-Lo\'{e}ve transform).
They demonstrated that SDSS galaxy spectra can be described by a small number of 
eigenspectra: more than 99\% of the  galaxies are found on a two-dimensional locus
in the space spanned by the ratios of the first three eigencoefficients. Even the 
most extreme emission-line spectra can be described to within the measurement noise 
with only 8 eigenspectra. An efficient {\it single-parameter} classification scheme 
is proposed, and it suggests three major galaxy classes. Another remarkable result of 
their analysis is a strong correlation between various spectral lines, encoded in, 
and readily discernible from the eigenspectra. 

It is not yet known what the results of Yip et al. imply for the detailed distribution 
of galaxies in color space, but it is obvious that a high degree of structure
should exist. Also, a strong correlation is expected between colors and various 
spectroscopic parameters, such as the strength of the $H_\delta$ line and the 4000~\AA\ break,
that were recently utilized by Kauffmann et al. (2003a) to estimate stellar masses and 
the the dust content for SDSS galaxies. 

The main goal of this paper is to investigate whether the intrinsic simplicity of galaxy 
spectra implied by the results of Yip et al. can be reproduced in rest-frame color-color 
diagrams, and whether the position of a galaxy in these diagrams is related to parameters 
discussed by Kauffmann et al. (2003a). We present an analysis of correlations between colors 
and various spectroscopic diagnostic parameters, including model-dependent estimates of 
stellar masses and dust content. We utilize the ``rest-frame'' Str\"omgren photometric system 
that measures the continuum spectral slope of galaxies in the rest-frame 3200--5800~\AA\ 
wavelength range. In principle, any set of bandpasses could be used to synthesize the colors.
An additional motivation for synthesizing galaxy colors in the rest-frame Str\"omgren 
photometric system is a suggestion that it is an efficient tool for studying faint cluster 
galaxies and low surface-brightness objects without performing time-consuming spectral 
observations\footnote{Observations in the rest-frame Str\"omgren photometric system use 
filters adjusted (in hardware) to the known cluster redshift (say, from the brightest 
cluster galaxy).} (Fiala, Rakos \& Stockton 1986). The unprecedented number of high-quality 
SDSS spectra can be used to analyze in detail the advantages and limitations of that method,
and relate galaxy colors in the Str\"omgren system to parameters commonly measured using 
SDSS data.

We describe the synthesis of rest-frame Str\"omgren magnitudes from SDSS
spectra in Section 2. In Section 3 we analyze the distribution of galaxies
in the resulting color-color diagram, and interpret it using the Bruzual~\& Charlot
(1993, 2003) population synthesis models. In Section 4 we demonstrate and study in 
detail correlations between rest-frame Str\"omgren colors and various diagnostic 
spectroscopic parameters. In Section 5 we correlate the rest-frame Str\"omgren
colors with the spectral eigencoefficients. In Section 6 we study the
properties of the galaxies at sufficiently large redshifts ($\gtrsim0.18$) to have
reliable synthesized \U~magnitudes. We discuss and summarize our results in Section 7.

\begin{figure} 
\centering
\includegraphics[bb=20 160 532 629, width=\columnwidth]{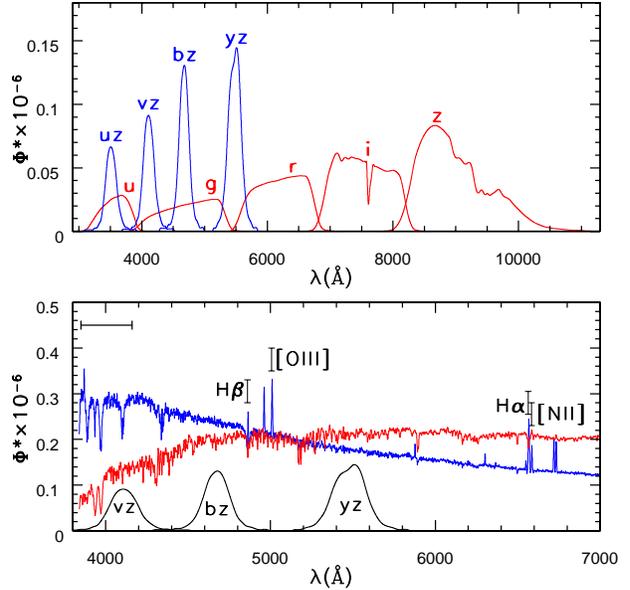}
\caption{The top panel shows renormalized (see eq.~\ref{phi}) filter
transmission curves for the SDSS photometric system (ugriz), and for the 
Str\"omgren photometric system. The bottom panel emphasizes the 3800-7000~\AA\
region. The two spectra are typical for blue and red galaxies, and 
the four labeled emission lines are used to separate star-forming from 
AGN galaxies.  The horizontal bar in the top left corner marks the 
wavelength region used in the analysis by Kauffmann et al. (2003).}
\label{fltsplot}
\end{figure}


\section{ The Synthesis of Rest-frame Str\"{o}mgren Photometry }

\subsection{The Str\"{o}mgren Photometric System}

The Str\"{o}mgren {\em uvby} narrow-band (\about 200~\AA) filter system was designed 
for measuring the temperature, chemical composition and surface gravity of stars, 
without resorting to spectroscopy (Str\"omgren 1966, for a recent compilation 
of over 100,000 measurements for $\sim$63,000 stars see Hauck \& Mermilliod 1998). 
The  bandpasses bracket the
4000~\AA\ break and cover three regions in the 3200-5800~\AA\ spectral region, which
makes it a powerful tool for investigating stellar populations in star clusters, or 
composite systems such as galaxies.

The filter system used here (\U, \V, \B, \Y; see next Section for details) is a 
somewhat modified Str\"{o}mgren system,
such that the filters are slightly narrower, and the filter response
curves are more symmetric, than in the original system (Odell et al. 2002). 
The system characteristics are (see Fig.~\ref{fltsplot}):
\begin{itemize}
  \item 
The \U\ filter ($\lambda_{eff}=3500$~\AA) is shortward from the Balmer discontinuity 
($4000\,\AA$ break), and provides a measure of hot stars due to recent star formation.
  \item 
The \V\  filter ($\lambda_{eff}=4100$~\AA) extends bluewards from $4600\,\AA$, but 
redwards from the Balmer discontinuity. This wavelength region is strongly influenced 
by metal absorption lines, particularly for spectral classes F and M which typically
dominate the galaxy light. The $uz-vz$ color is a good measure of the $4000$~\AA~break.
  \item 
The \B\ ($\lambda_{eff}=4675$~\AA) and \Y\ ($\lambda_{eff}=5500$~\AA) filters extend
redwards from 4600~$\AA$, where the influence of absorption lines is small.
The $bz-yz$ color is thus a good measure of the temperature color index, practically
free of metallicity and surface gravity effects. For old stellar populations, the \VY\ and 
\BY\  color indices presumably serve as age and metallicity indicators. 
\end{itemize}

\subsection{The Rest-frame Str\"{o}mgren Photometric System}
 
The rest-frame Str\"{o}mgren photometric system utilizes filters whose
bandpasses are redshifted and stretched to correspond to the Str\"{o}mgren bands in
the rest-frame. This method, introduced by Fiala, Rakos \& Stockton (1986),
alleviates the need for $K-$corrections, and thus allows robust color comparison 
for galaxies at different redshifts. For example, Rakos \& Schombert (1995) studied 
the color evolution of 17 clusters of galaxies spanning the redshift range from 
$0.2$ to $0.9$. They demonstrated that the colors of the red population are consistent 
with passive evolution models for a single stellar population which formed
around $z=5$. On the other hand, the fraction of blue cluster galaxies 
dramatically increases from $20\%$ at $z=0.4$ to $80\%$ at $z=0.9$.
The rest-frame Str\"{o}mgren photometric system also provides a good method for 
detecting cluster members, and enables an efficient spectrophotometric
classification of galaxies (Rakos, Dominis and Steindling 2001; Rakos \& Schombert 2004; 
and references therein).


\begin{figure}
\centering
\includegraphics[bb=22 70 580 668, width=\columnwidth]{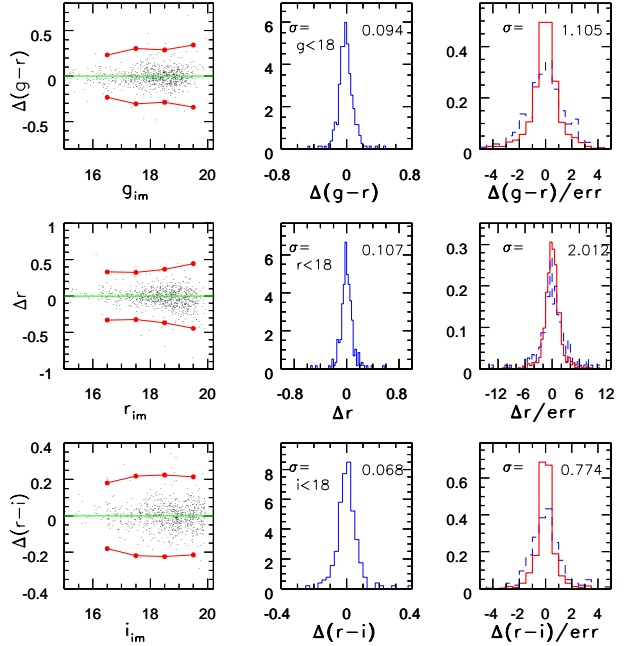}
\caption{The comparison of imaging and synthesized SDSS magnitudes (colors) for stars, with the latter
tied to imaging point-spread-function magnitudes.
The first column displays the color or magnitude difference as a function of magnitude;
the large symbols connected by lines show the $\pm 3 \sigma$ envelope. The distribution 
of color or magnitude differences at the bright end is displayed in the middle column.
The last column shows the distribution of color or magnitude differences normalized
by the expected errors at the bright end (solid line), and for the full sample 
(dashed line).}
\label{dmStars}
\end{figure}



\begin{figure}
\centering
\includegraphics[bb=72 124 540 618, width=\columnwidth]{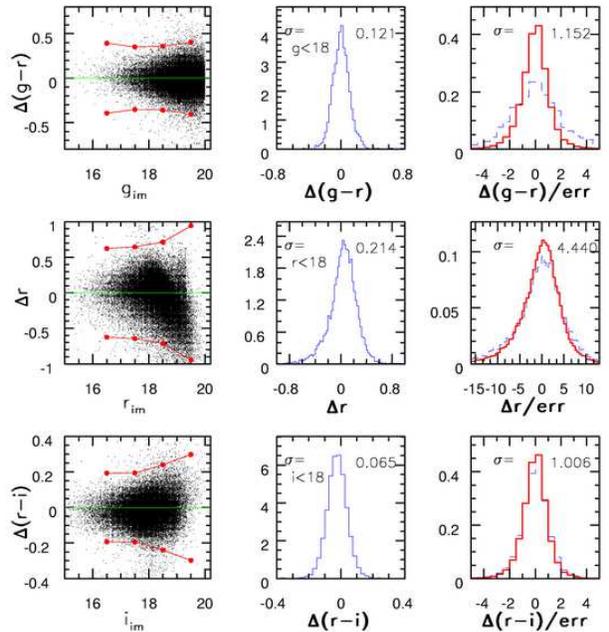}
\caption{Analogous to Fig.~\ref{dmStars}, except that the results are shown for galaxies,
and the spectrophotometric magnitudes are tied to imaging model magnitudes.}
\label{dmGalaxies}
\end{figure}

\subsection{The Synthetic Str\"omgren Photometry from SDSS spectra  }

The synthetic Str\"omgren magnitudes are synthesized on the AB 
system\footnote{Note that all the published work by Rakos and collaborators, 
who pioneered the redshifted Str\"omgren photometric system, is expressed on the
Vega system.} (Oke \& Gunn  1983). An AB magnitude, $m$, in a bandpass $S(\nu)$, is defined by

\begin{equation}
\label{fukugita}
m=-2.5\log_{10}\frac{\int{d(\log_{10}\nu) f_{\nu}(\nu) S(\nu)}}{\int{d(\log_{10}\nu)S(\nu)}}-48.6,
\end{equation}

where $f_{\nu}$ is the specific flux per unit frequency (for more details
see Fukugita et al. 1996). The zero-point (48.6) is chosen such that an object with 
a specific flux of 3631 Jy has $m=0$ (i.e. an object with $f_{\nu}$=const. has
an AB magnitude equal to the Johnson $V$ magnitude at all wavelengths).
The function $S(\nu)$ includes the response of the detector, the
transmissivity of the filter, the telescope, and the atmosphere of the
Earth at some representative air-mass. It is convenient to express the
above equation in terms of the specific flux per unit wavelength (the
form used to report SDSS spectra, Stoughton et al. 2002) as

\begin{equation}
\label{mag2}
   m=-2.5\log_{10}\int{\phi^\ast(\lambda)f_\lambda(\lambda)d\lambda},
\end{equation}

where we introduced a quantity, hereafter called the {\em normalized filter}:

\begin{equation}
\label{phi}
\phi^\ast (\lambda) = \frac{10^{19.44} \lambda S(\lambda)}{c \int{\lambda^{-1}S(\lambda) d\lambda}}
\end{equation}

Computing {\em m} is, therefore, reduced to a simple integral of the SDSS spectrum 
(with the spectrum given in 10$^{-17}$ erg/cm$^2$/s/\AA, and $\lambda$ in \AA) with a 
{\em pre-computed} normalized filter. The normalized filter curves are shown in 
Fig.~\ref{fltsplot}.

SDSS spectra are obtained using a plate with 640 spectroscopic fibers (diameter 3",
corresponding to \about6 kpc at the redshift of 0.1), which are fed to two independently 
calibrated spectrographs (Schlegel et al., in prep). Spectra are corrected for the
Galactic interstellar dust reddening during calibration. As shown by Vanden Berk et al. (2002), 
the accuracy of synthesized magnitudes for point sources can be improved by requiring that 
the mean offset between the imaging ($g$, $r$, $i$) and synthesized magnitudes is zero for 
each spectrograph (this information is not available when calibrating spectra). We 
confirmed their result for point sources, and found that it also improves the 
synthesized photometry for galaxies (as inferred from the width of the galaxy color
locus, discussed below). The synthesized photometry for galaxies can be further
improved by using galaxies (instead of point sources) to correct for the mean offsets 
between the imaging (we use model magnitudes, see Stoughton et al. 2002
for details) and synthesized magnitudes. Hereafter, we adopt the latter method. 
When correcting synthesized Str\"omgren magnitudes, we linearly interpolate (as a function 
of wavelength) the $g$ and $r$ band corrections.

\subsection{The Tests of Synthesized Magnitudes}

We synthesized SDSS magnitudes ($g$, $r$, $i$) and rest-frame Str\"omgren magnitudes 
(\U, \V, \B, \Y)\footnote{ 
   We discuss the implications of the \U~magnitude separately in
   Section~\ref{sec:uz} since it can only be synthesized 
   for galaxies with redshifts beyond $\sim$0.18.   } 
for 99,088 unique ``main'' ($r_{Pet}<17.77$) galaxies from SDSS Data 
Release 1 (Abazajian et al. 2003), as well as for stars from the same spectroscopic plates. 
The accuracy of synthesized magnitudes (relative to imaging magnitudes, which are accurate 
to 0.02~mag, Ivezi\'{c} et al. 2003) for a subsample of stars is summarized in 
Fig.~\ref{dmStars}. The colors are reproduced with the expected accuracy (\about0.08~mag), 
while there are systematic errors in the overall magnitude scale at the level of 0.10~mag 
(root-mean-square scatter). These results are in agreement with Vanden Berk et al. (2004).

Fig.~\ref{dmGalaxies} summarizes the obtained accuracy for galaxies. Again, the colors
are reproduced with the expected accuracy, while systematic errors in the overall magnitude 
scale are \about0.2~mag. Note that using galaxies (instead of point sources) to correct for 
the offset between imaging and synthesized magnitudes introduces a systematic trend in the
$\Delta r$ distribution at the faint magnitude end. Nonetheless, as evident from the top
and bottom panels, this does not significantly affect the synthetic colors as the overall 
trend nearly cancels out. In addition, the distribution of color-differences normalized by the
expected errors (last column in Fig.~\ref{dmGalaxies}, top and bottom panels) shows that the 
errors are estimated correctly (the width of the distributions is \about1).
Therefore, we do not expect this to affect the synthetic photometry significantly.
The errors in rest-frame Str\"omgren colors are expected to be
smaller than in $g-r$ and $r-i$ colors because the relevant wavelength range
is shorter (and thus less sensitive to large-scale errors in the spectrophotometric 
calibration of SDSS spectra). We estimated errors in $vz-yz$ and $bz-yz$ colors by 
comparing measurements for 1253 galaxies observed twice: the scatter (root-mean-square 
determined from the interquartile range) for the $vz-yz$ color is 0.09~mag,
and for $bz-yz$ color 0.05~mag. 
We tested for plate-to-plate systematics by analyzing the position of the red peak
in the color distribution of galaxies (see the next Section). This position shows a 
scatter of $\sim0.02$~mag for $vz-yz$ color, and $\sim0.01$~mag for $bz-yz$ color.
We conclude that the 
synthesized colors are sufficiently accurate for a detailed investigation  
of rest-frame colors. This is the first time that rest-frame optical colors with 
an accuracy of $<$0.1~mag are available for a sample of \about100,000 galaxies.

\vspace{0.8cm}

\section{ The Colors of Galaxies in the Rest-frame Str\"omgren system }

In this Section we analyze the distribution of galaxies in the rest-frame Str\"omgren
color-color diagram, discuss possible systematic errors and interpret the
colors using model spectra.
 
\subsection{ The Galaxy Locus and Principal Axes } 
\label{galLocus}


\begin{figure}
\centering
\includegraphics[bb=35 169 577 622, width=\columnwidth]{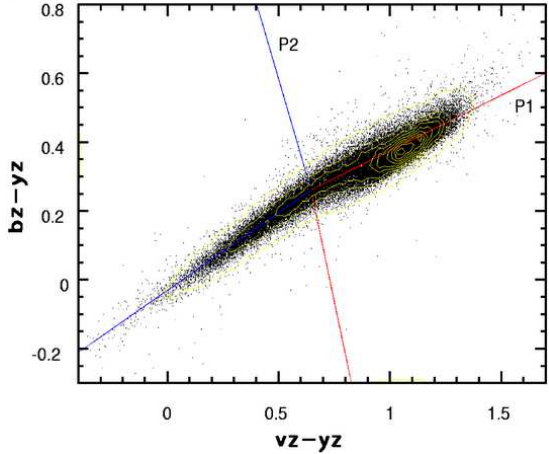}
\caption{The distribution of 99,088 SDSS ``main'' galaxies in the rest-frame 
Str\"omgren color-color diagram. Each galaxy is represented by a dot, and 
the overall distribution is outlined by linearly spaced isopleths. The principal 
axes P1 and P2 are defined for the two regions separated by $vz-yz=0.646$.}
\label{galcorr}
\end{figure}


\begin{figure}
\centering
\includegraphics[bb=40 270 572 500, width=\columnwidth]{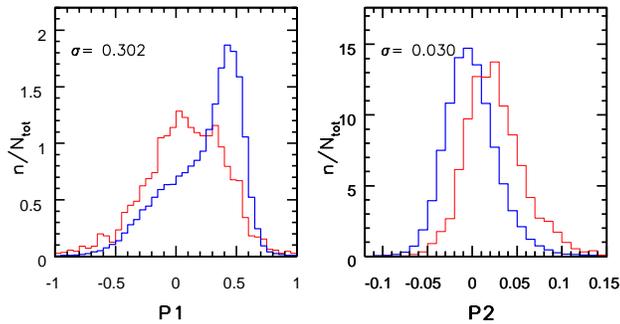}
\caption{The color probability distributions of the principal rest-frame Str\"omgren colors for SDSS 
``main'' galaxies (note different scales for x axis). The thick lines show
the distributions for the whole sample, and thin lines for the subsample of
galaxies detected by IRAS. The $\sigma$ values shown in each panel are
root-mean-square scatter (determined from the interquartile range) for the
whole sample.} 
\label{P1P2hist}
\end{figure}

The distribution of 99,088 galaxies in the rest-frame Str\"omgren color-color diagram
is shown in Fig.~\ref{galcorr}. The distribution resembles a remarkably narrow locus.
To quantify its width, we define a set of principal axes ($P1,P2$) where $P1$ 
measures the position along the locus, and $P2$ perpendicular to it. Since the
locus is not perfectly straight, we separately fit the blue ($vz-yz<0.6$) and red 
($vz-yz>0.7$) ends. As the boundary between the two ($P1,P2$) definitions we 
choose $vz-yz=0.646$, where the two fits intersect. We obtain for the blue
side

\eq{
\label{Paxes1}
   P1 = \phantom{X}0.911\,(c_1-0.646) + 0.412\,(c_2-0.261)
}
\eq{
   P2 = -0.412\,(c_1-0.646) + 0.911\,(c_2-0.261),
}
and 
\eq{
   P1 = \phantom{X}0.952\,(c_1-0.646) + 0.307\,(c_2-0.261) \\
}

\eq{
\label{Paxes2}
   P2 = -0.307\,(c_1-0.646) + 0.952\,(c_2-0.261),
}

for the red side, with $c_1=vz-yz$ and $c_2=bz-yz$. The coefficients are normalized such 
that the errors in $P1$ and $P2$ are the same as the errors in $vz-yz$ and $bz-yz$, in the 
limit when the latter two are the same. It is noteworthy that the $P2$ axis is 
nearly parallel to the color index $m_1 = (vz-yz)-2\,(bz-yz)$, used as a metallicity
estimator for single stars (e.g. Twarog 1980).

The distributions of $P1$ and $P2$ colors are shown in Fig.~\ref{P1P2hist}
by thick lines. The width of the locus in the P2 direction is surprisingly 
small\footnote{It may be surprising that the width of the P2 color distribution
is smaller than the errors of the colors used to synthesize it. This is due 
to covariances in the $vz-yz$ and $bz-yz$ errors.}
(0.03~mag, and 0.32~mag in the P1 direction). Indeed, the rest-frame colors for 
a large number of galaxies have not 
been measured to date with such an accuracy. This small value both testifies that the 
errors in synthesized colors are small, and that {\it the slope of galaxy spectra in 
the 4000--5800~\AA\ wavelength range is a nearly one-parameter family}.

\subsection{ Fiber Aperture Effects on P1 \& P2}

\label{APeff}

SDSS spectra are obtained with a fixed 3 arcsec diameter fiber positioned as
close as possible to the center of the galaxy. Since the fiber aperture 
may not include the entire galaxy, it is possible that colors derived
from SDSS spectra are biased red by the bulge contribution. This bias
should be the largest for intermediate- and late-type galaxies where 
the fiber aperture may miss a significant fraction of the outer disk and
its star formation. While this effect was discussed in 
detail elsewhere (e.g. Kauffmann et al. 2003a and Kewley et al. 2005),
here we explore its impact on the P1 and P2 colors.

The fiber aperture effect will tend to make the P1 color redder for 
galaxies with large angular size because bulges are generally redder than disks. 
The expected effect for the P2 color is not so obvious. We argue 
below (see Section~\ref{irasSec}) that the P2 color is related to dust,
and if this is true, the expected effect is for the P2 color to become
bluer for galaxies with large angular size. 

\begin{figure}
\centering
\includegraphics[bb= 85 177 522 539, width=\columnwidth]{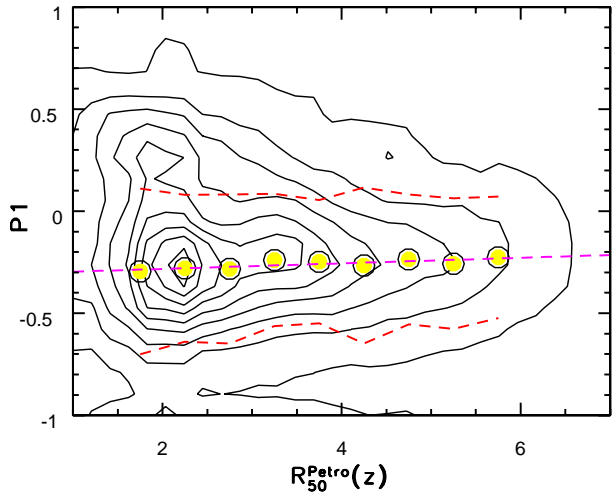}
\caption{A test for aperture effects, due to 3 arcsec SDSS spectroscopic
fiber diameter, on the P1 color. The linearly spaced contours show the distribution
of galaxies with absolute Petrosian magnitude in the $r$ band between -19 and -18
in the P1 color vs. Petrosian 50\% radius in the $z$ band (an estimator of the 
galaxy's size) diagram. The aperture effect is expected to be the
strongest for blue galaxies with P1$<$0. The large circles show median P1 colors 
for bins of the Petrosian radius in the $z$ band, computed for galaxies
with P1$<$0. The middle line is a best straight line fit to these medians, 
and the other two lines illustrate 2$\sigma$ deviation around the median. 
The slope of the best fit line is 0.015~mag/arcsec.}
\label{APeffP1}
\end{figure}

\begin{figure}
\centering
\includegraphics[bb= 85 177 522 539, width=\columnwidth]{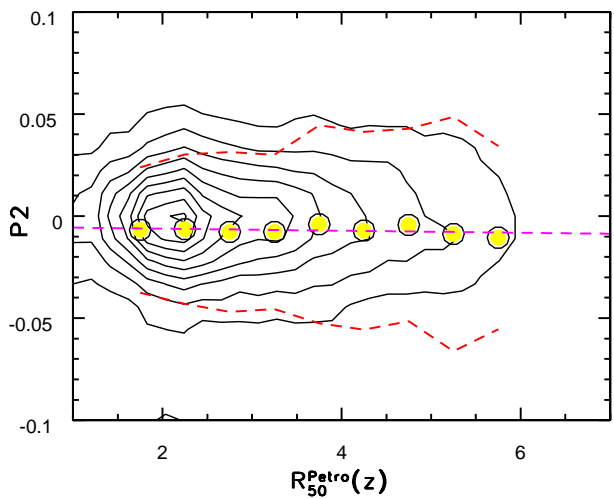}
\caption{Analogous to Fig.~\ref{APeffP1}, except for the P2 color and only for
galaxies with P1$<$0. The slope of the best fit line is -0.0005~mag/arcsec. 
}
\label{APeffP2}
\end{figure}

Figures~\ref{APeffP1}~and~\ref{APeffP2} show the dependence of the P1 and
P2 colors on the Petrosian 50\% radius in the $z$ band\footnote{For more
details about the Petrosian radii see Stoughton et al. 2002, and Strauss 
et al. 2002.} for galaxies with the $r$ band absolute magnitudes in the 
range -19 to -18. This magnitude slice is selected because it is dominated
by blue galaxies where the expected effect is the strongest 
(see Fig.~\ref{MrP1}). As the Petrosian radius increases, the bulge 
contribution to the flux captured by the fiber also increases. 
As evident, small but measurable dependence of the P1 color on the 
Petrosian radius is indeed present. Given the observed range of the 
Petrosian radius, the expected bias in the P1 color is $\la$0.05~mag,
with the root-mean-square scatter contribution to the P1 measurement error
of $\sim$0.02~mag. Compared to the observed P1 range, this is a small 
effect. The  fiber aperture effect on the P2 color is entirely negligible.

\subsection{ The Stellar Population Model Colors }

\label{modelsSec}


\begin{figure}
\centering
\includegraphics[bb=20 60 592 699, width=\columnwidth]{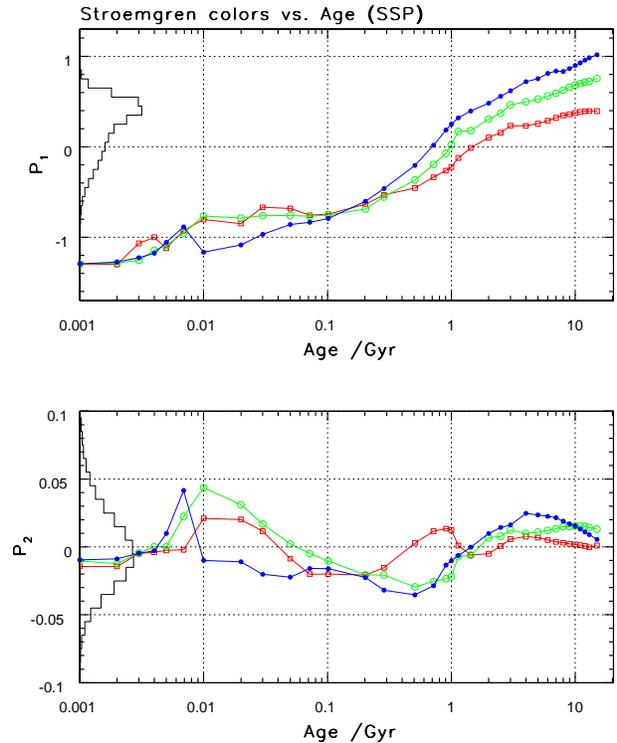}
\caption{The Bruzual~\& Charlot model predictions for the dependence of the principal 
rest-frame Str\"omgren colors on the age and metallicity ([Z]=0.004, squares; 0.02, 
circles; 0.05, dots) of single stellar populations. Note the different scales
for the two colors. The histograms on the y axes show the observed color
distributions.}
\label{BCmodels}
\end{figure}

The principal axes (P1,P2) are defined by the {\it observed} morphology of 
the distribution of galaxies in the rest-frame Str\"omgren color-color diagram. 
Here we relate these axes to the stellar age and metallicity using
the Bruzual~\& Charlot (1993, 2003) population synthesis models.

Using the same method as for observed spectra, we synthesize 
rest-frame Str\"omgren colors from model spectra. Fig.~\ref{BCmodels}
shows the model colors as a function of the age of a single stellar
population, and for three representative metalicities. A single stellar 
population corresponds to a single epoch of star formation, while real 
galaxies likely have more complex star formation histories. In addition,
for single stellar population models the star-formation rate history,
which is an important discriminating parameter (e.g. Kennicutt 1998),
is unavailable. Nevertheless, 
the color range spanned by a single stellar population provides a robust 
constraint on the possible color distribution for any star formation history
(because it brackets the possible color range).

The observed range of the $P1$ color can be explained by stellar populations 
with ages from 100 Myr to 10~Gyr. Note that $P1>0$ corresponds to populations 
older than \about1~Gyr, but the age estimate is very uncertain due to
the unknown metallicity.

Models provide a simple explanation for the 
observed small locus width: the $P2$ color is a color projection which is nearly 
independent of both age {\it and} metallicity\footnote{The $P2$ color, while much 
less sensitive to metallicity than the $P1$ color, does become redder with increasing 
metallicity for populations older than 1 Gyr. However, the effect is exceedingly 
small (0.03 mag per dex). For individual stars ($P2$ is nearly parallel to $m_1$, 
c.f. section \ref{galLocus}) this is large enough because their effective temperature 
is typically constrained much better than the star-formation history of a galaxy.}.
At any age, $P2 \about 0$ to within 0.02-0.03 mag, and for populations older than 
\about1~Gyr, $P2 \about 0.01$ to within 0.01~mag. That is, whatever is age or 
metallicity, $P2$ is always 0 to within a few hundreths of a magnitude. This is very 
different from $P1$ which varies by 2 magnitudes as a function of age, and up to 0.6 mag 
as a function of metallicity. {\it The ability of models to explain the mean value and 
the very narrow distribution of the $P2$ color is a considerable success because the 
principal axes $P1$ and $P2$ are defined solely by the morphology of the data distribution 
in the $bz-yz$ vs. $vz-yz$ color-color diagram.} 

The observed width of the $P2$ distribution is wider (0.03~mag) than the model 
prediction (\about0.01~mag). While this difference could be easily attributed
to the measurement scatter or to the simplicity of the models, we find that
the $P2$ color is correlated with the estimate of the galaxy dust content, 
$A_z$ (the effective dust extinction in the SDSS $z$ band), determined by Kauffmann 
et al. (2003a) and further discussed in Section \ref{KauffSec}. Hence, the larger 
scatter observed in the $P2$ color than predicted by the model is, at least partially, 
due to dust reddening (which is not included in the models).

It could be argued that the $P2-A_z$ correlation is simply due to the fact
that both quantities are determined from SDSS spectra. However, they are
determined using {\it different}, barely overlapping, spectral ranges. While 
Kauffmann et al. analysis utilized the 3850--4160~\AA\ spectral range to constrain 
their models (and then compared imaging $g-r$ and $r-i$ colors to model-predicted
colors to estimate dust reddening), $P2$ is determined from the 4000--5800~\AA\ 
wavelength range (see the bottom panel in Fig.~\ref{fltsplot}).

Another reasonable objection is that the $A_z$ estimates are, of course, model dependent 
and consequently may not be a good measure of the dust content. However, Obri\'{c} 
et al. (2006) demonstrate that galaxies detected by IRAS have systematically higher 
values of $A_z$. This finding strongly suggests that $A_z$ does provide an 
estimate of the dust content.

\subsection{   The Colors of IRAS-detected galaxies }
\label{irasSec}

Since the $P2-A_z$ correlation implies that $P2$ is also a measure of the galaxy dust 
content, galaxies detected by IRAS should have systematically higher values 
of $P2$. Using the same SDSS-IRAS galaxy sample discussed by Obri\'{c} et al.
(\about2200 probable matches for the sample of 99,088 galaxies discussed here), 
we compared the $P1$ and $P2$ distributions for IRAS-detected galaxies to those
for the full sample. As evident from Fig.~\ref{P1P2hist}, galaxies detected by IRAS 
have systematically higher $P2$ values.

\begin{figure}
\centering
\includegraphics[bb=30 80 552 720, width=\columnwidth]{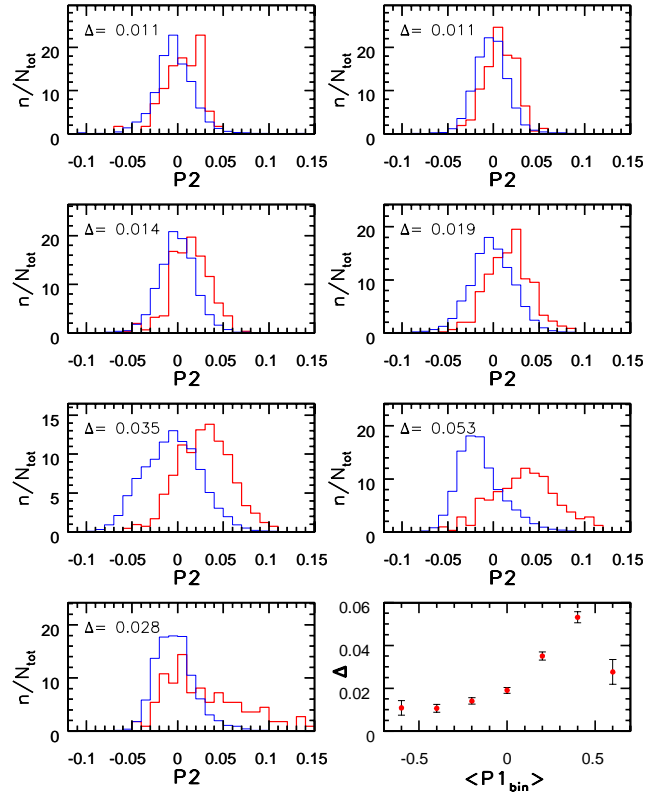}
\caption{The comparison of $P2$ distributions for all galaxies and for the
subsample detected by IRAS, for 0.2~mag wide bins in $P1$, in the range
-$0.7$ to 0.7 (row by row, from top left to bottom right). The thin lines show 
the distributions for the whole sample, and the thick lines for the IRAS-detected
subsample. The $\Delta$ values shown in each panel are the difference between the 
medians for the two distributions, and are also shown as a function of $P1$ in the 
bottom right panel.}
\label{P2panels}
\end{figure}

The linear definition of the principal axes $P1$ and $P2$ (eqs.~[\ref{Paxes1}]--[\ref{Paxes2}])
is not a perfect description of the galaxy locus shown in Fig.~\ref{galcorr} because the
locus is curved. Hence, the $P2$ distribution slightly changes along the locus.
Since IRAS-detected galaxies are biased towards blue galaxies, it could be 
that the difference in $P2$ distributions for IRAS-detected and all SDSS galaxies
shown in Fig.~\ref{P1P2hist} is caused by the variation of the $P2$ distribution
along the locus. Fig.~\ref{P2panels} demonstrates that this is not the case:
for each segment of the locus defined by a narrow bin of $P1$, IRAS-detected
galaxies have systematically higher $P2$ colors. This offset strongly suggests
that $P2$ is a measure of the galaxy dust content.

The shift of the $P2$ distributions between SDSS-IRAS and all SDSS galaxies is 
about the same as the $P2$ distribution width, so $P2$ is a rather noisy measurement
of the dust content. Nevertheless, we demonstrate below that it is sufficiently 
efficient to study differences in the dust content between various galaxy populations
separated using emission lines.


\section{ The Correlations Between the Rest-frame Str\"omgren Colors and
                             Spectral Parameters }

In the previous Section we have shown the connection between the two principal rest-frame 
Str\"omgren colors and galaxy physical parameters: $P1$ depends, in a degenerate way,
on stellar age and metallicity, while $P2$ is correlated with the dust content,
and independent of age and metallicity.
In this Section we further analyze correlations between Str\"omgren colors and 
various spectral parameters. We show that the position of an emission line galaxy
in a diagram commonly used to separate star-forming from AGN galaxies is well
correlated with the $P1$ color, that $P1$ and $P2$ colors are correlated with
the parameters discussed by Kauffmann et al. (2003ab), and that the rest-frame $r-i$ 
color can be predicted to better than 0.05~mag from the $P1$ color.  All these
correlations suggest that galaxy spectra are simple and well-defined 
superpositions of stellar spectra.

\subsection{ The Baldwin-Phillips-Terlevich Diagram }


\begin{figure}
\centering
\includegraphics[bb=92 72 520 720, width=\columnwidth]{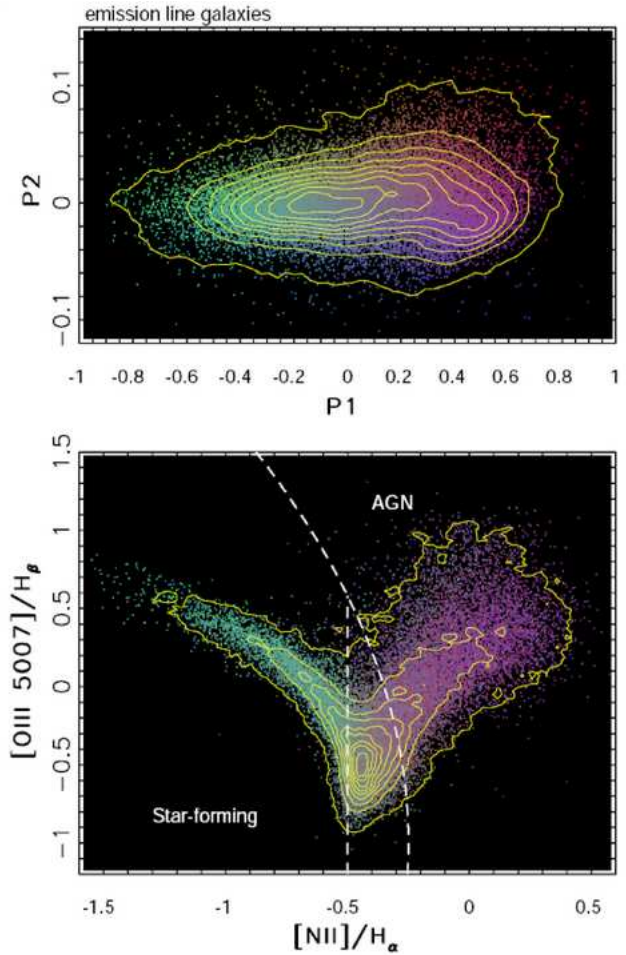}
\caption{The top panel shows the distribution of SDSS ``main'' galaxies
in a diagram constructed with the principal rest-frame Str\"omgren colors.
Each dot corresponds to one galaxy; their distribution is outlined by 
linearly spaced isopleths. The color code is determined by the position
in this diagram, and is used in subsequent figures to visualize correlations
of other quantities with $P1$ and $P2$. The bottom panel shows the distribution
of galaxies in the Baldwin-Phillips-Terlevich diagram, constructed with
emission-line strength ratios. The dashed lines separate the regions populated
by star-forming and AGN galaxies. They are motivated by the modeling results
from Kewley et al. (2005) and are fine-tuned to isolate the peak of the 
observed distribution. For galaxies in the wedge-shaped regions
the classification is uncertain. The dots are colored using $P1$ and $P2$;
note the strong correlation between the $P1$ color and $[NII]/H\alpha$ ratio, 
although these parameters are determined in different wavelength ranges, 
and measure the continuum slope and line strengths, respectively.}
\label{BPT}
\end{figure}


\begin{figure}
\centering
\includegraphics[bb=70 200 522 509,width=\columnwidth]{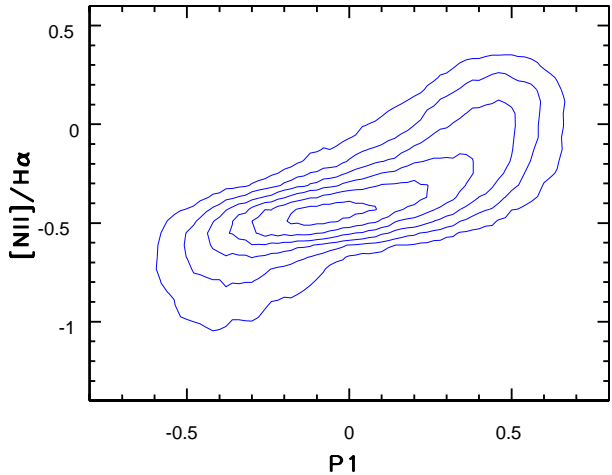}
\caption{The correlation between the first principal axis in the 
rest-frame Str\"omgren color-color diagram, and the $[NII]/H\alpha$
line strength ratio. The distribution of all emission-line galaxies  
in the sample is shown by linearly spaced isopleths.}
\label{P1NIIHa}
\end{figure}


\begin{figure}
\centering
\includegraphics[bb=50 90 502 629, width=\columnwidth]{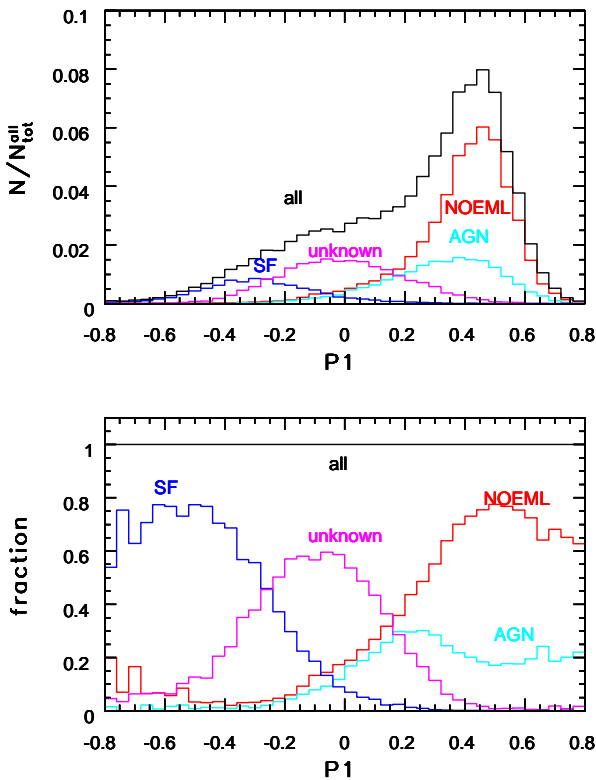}
\caption{The distributions of $P1$, the first principal axis in the 
rest-frame Str\"omgren color-color diagram, for subsamples of galaxies 
separated using emission line strengths: galaxies without emission 
lines (NOEML), star-forming galaxies (SF), AGNs, and emission-line 
galaxies which cannot be reliably classified (unknown). The top panel
displays counts, and the bottom panel displays the fraction of each
subsample in the whole sample.
}
\label{P1classFracs}
\end{figure}


\begin{figure}
\centering
\includegraphics[bb=50 90 502 679, width=\columnwidth]{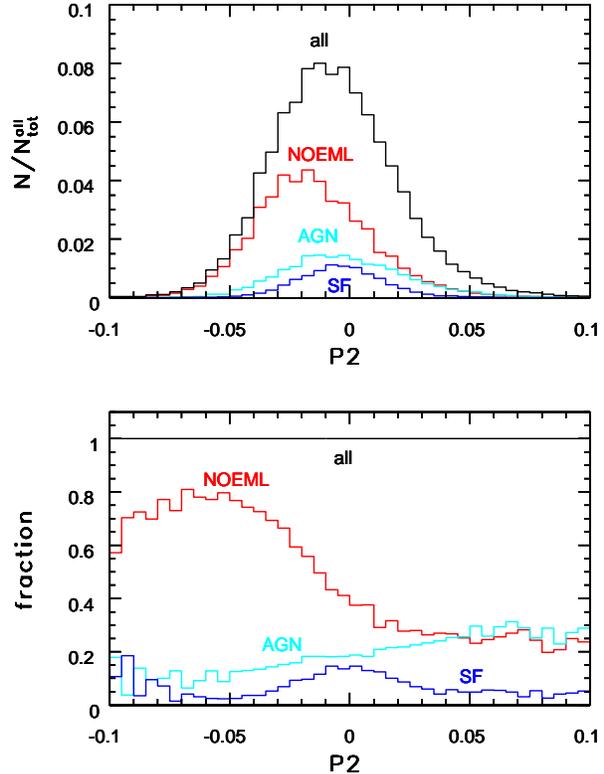}
\caption{The distributions of $P2$, the second principal axis in the 
rest-frame Str\"omgren color-color diagram, for subsamples of galaxies 
separated using emission line strengths: galaxies without emission 
lines (NOEML), star-forming galaxies (SF) and AGNs. The top panel
displays counts, and the bottom panel displays the fraction of each
subsample in the whole sample (emission-line galaxies which cannot be 
reliably classified make up the missing fraction).
}
\label{P2classFracs}
\end{figure}

The BPT diagram (Baldwin, Phillips and Terlevich, 1981) is a standard method 
for classifying emission-line galaxies as star-forming and AGN galaxies.
With the advent of SDSS data, it is now possible to construct the BPT diagram
for an unprecedented number of galaxies (e.g. Ivezi\'{c} et al. 2002, Kauffmann 
et al. 2003c, Hao et al. 2005ab, Heckman et al. 2004). Obri\'{c} et al. (2006)
have shown that the presence of emission lines in a galaxy's spectrum is well 
correlated with the $u-r$ color and the concentration index determined {\it solely} from 
the imaging data. Galaxies without emission lines tend to have larger concentration 
indices and redder $u-r$ colors than galaxies with emission lines. Furthermore, 
the distribution of emission-line galaxies in the BPT diagram is also correlated 
with the $u-r$ color and concentration index. Galaxies classified as star-forming
have predominantly blue $u-r$ colors and small concentration indices, while AGN 
galaxies have redder $u-r$ colors and large concentration indices.

Here we investigate the correlation between the strength of the emission lines used 
to construct the BPT diagram, and rest-frame Str\"omgren colors. To classify a 
galaxy as an emission-line galaxy, we require a $3\sigma$ significant detection 
of $H_\alpha$, $H_\beta$, $[NII]$ and $[OIII\,5007]$  lines (the line strengths
are determined as described by Kauffmann et al. 2003c). The bottom panel in 
Fig.~\ref{BPT} shows the distribution of 43,281 emission-line galaxies, from 
the sample discussed here, in the BPT diagram. To visualize the correlation with 
$P1$ and $P2$ colors, the dots are two-dimensionally color coded according to their 
position in the $P1$-$P2$ color-color diagram shown in the top panel.
{\it There is a strong correspondence between the position of a galaxy in the
BPT diagram and its $P1$ color.} Galaxies in the ``star-forming branch'' with small 
$[NII]/H_\alpha$ ratios, for a given $[OIII\,5007]/H_\beta$ ratio, have predominantly 
blue $P1$ colors, while AGN galaxies have redder $P1$ colors. This correlation 
is further illustrated in Fig.~\ref{P1NIIHa}. The correlation with the $P2$ color is 
not discernible.

In the subsequent analysis, we separate emission-line galaxies in three
groups according the their position in the BPT diagram: AGNs, star-forming, 
and ``unknown''. The adopted separation boundaries are shown by the dashed lines.
The last category is found at the joint of the two branches, and it is not obvious 
from the displayed data to which class these galaxies belong. 
Fig.~\ref{P1classFracs} compares the $P1$ distributions of these classes. 
The majority of galaxies without emission lines and AGN galaxies have $P1>0$,
while star-forming galaxies have $P1<0$. More than \about2/3 of galaxies with 
$P1>0.3$ are galaxies without emission lines, and more than \about2/3 of 
galaxies with $P1<-0.4$ are star-forming galaxies. At least 20\% of the galaxies
with $P1>0$ harbor an AGN. The fraction of AGN galaxies drops to essentially
zero by $P1\sim-0.3$ (or by $P1\sim-0.5$ if {\it all} unclassified galaxies 
are AGNs). This dramatic absence of AGNs among the bluest galaxies appears
not to be a selection effect, as discussed by Kauffmann et al. (2003c, Section 4.1).

The comparison of $P2$ distributions for the three subclasses of galaxies
is shown in Fig.~\ref{P2classFracs}. As expected, AGNs and star-forming 
galaxies have redder $P2$ distributions than galaxies without emission lines.

\subsection{         The $H_\delta$ -- $D_n(4000)$ Locus }

\label{KauffSec}

\begin{figure}
\centering
\includegraphics[bb=85 36 526 757, width=\columnwidth]{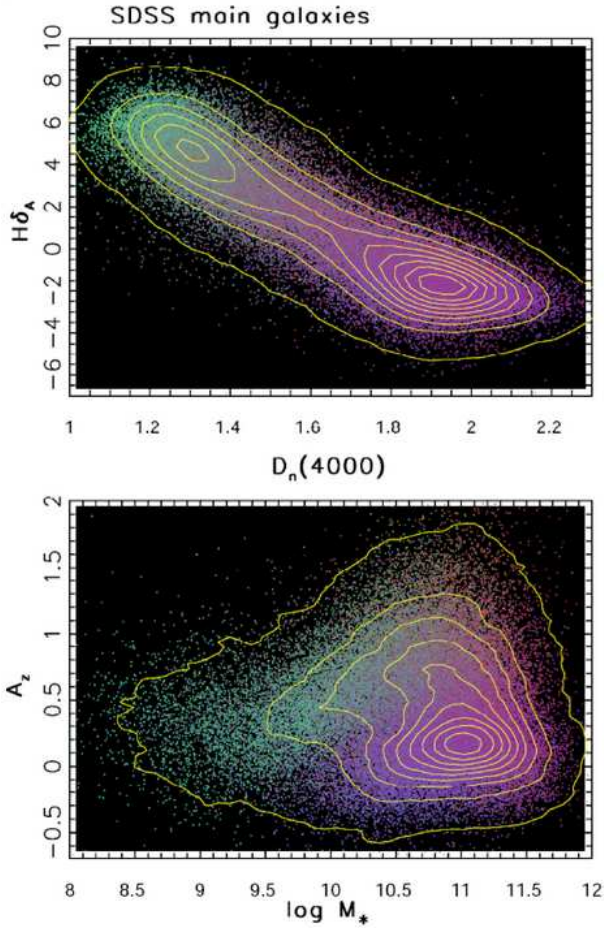}
\caption{The top panel shows the distribution of SDSS galaxies in the plane 
spanned by the strength of the $H_\delta$ line and the 4000~\AA\ break ($D_n(4000)$).
Each galaxy is represented by a point color-coded according to its position 
in the $P1$-$P2$ color-color diagram, shown in the top panel in Fig.~\ref{BPT}. 
Note the strong color gradient indicating a correlation between the position 
of a galaxy along this locus and its $P1$ color. The overall, strongly bimodal, 
distribution is outlined by linearly spaced isopleths. The bottom panel shows
the distribution of the same sample in the dust content ($A_z$) vs. stellar 
mass ($M_\ast$) diagram. $A_z$ and $M_\ast$ are model-dependent estimates
derived from the measured $H_\delta$, $D_n(4000)$ and imaging $g-r$ and $r-i$ 
colors. The color gradients indicate good correlations between the positions
in the $A_z$-$M_\ast$ and $P1$-$P2$ planes.}
\label{etaAll}
\end{figure}


\begin{figure}
\centering
\includegraphics[bb=80 57 532 739, width=\columnwidth]{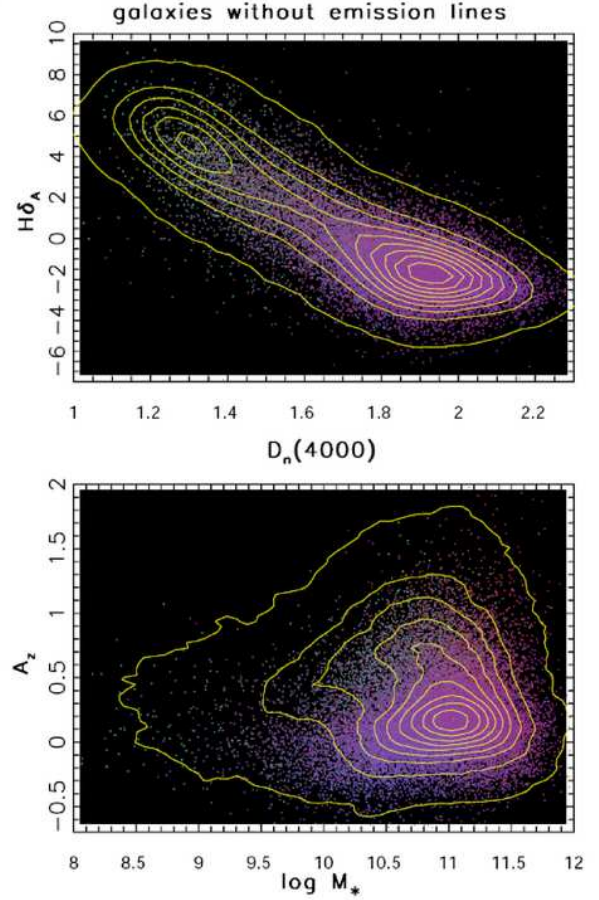}
\caption{Analogous to Fig.~\ref{etaAll}, except that only galaxies without
emission lines are shown (contours correspond to the full sample).}
\label{etaNOEML}
\end{figure}


\begin{figure}
\centering
\includegraphics[bb=122 124 500 688, width=\columnwidth]{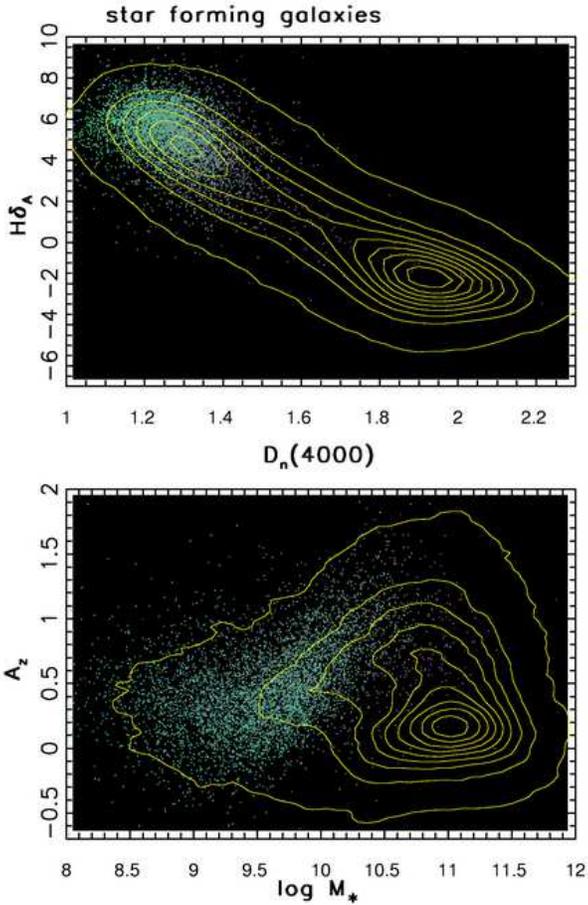}
\caption{Analogous to Fig.~\ref{etaAll}, except that only star-forming galaxies
are shown (contours correspond to the full sample).
}
\label{etaSF}
\end{figure}


\begin{figure}
\centering
\includegraphics[bb=122 124 500 688, width=\columnwidth]{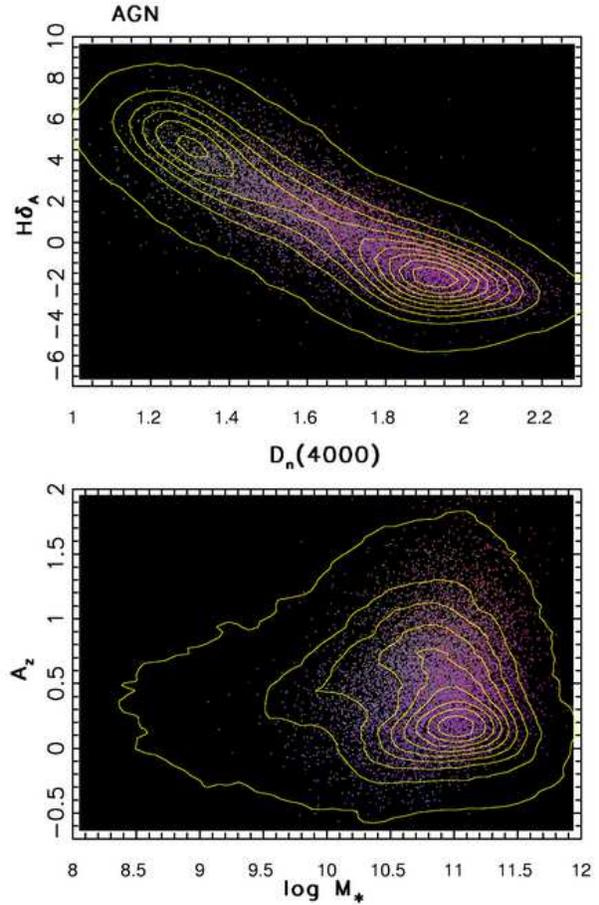}
\caption{Analogous to Fig.~\ref{etaAll}, except that only AGN galaxies
are shown (contours correspond to the full sample).}
\label{etaAGN}
\end{figure}


\begin{figure}
\centering
\includegraphics[bb=80 57 532 759, width=\columnwidth]{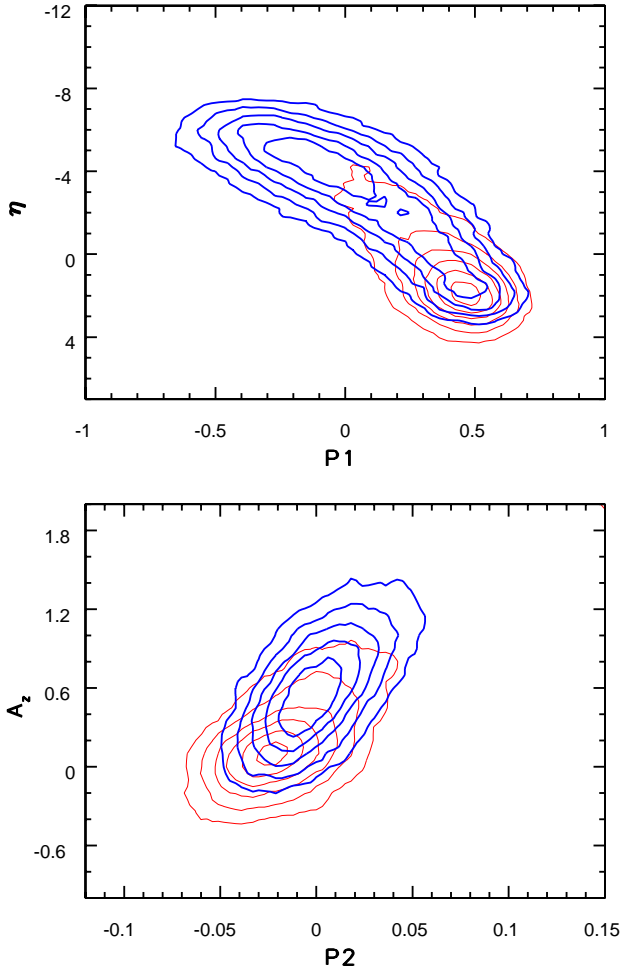}
\caption{The top panel shows the correlation between the position of a galaxy on the locus
in the $H_\delta$-$D_n(4000)$ plane ($\eta$) and the $P1$ color. The thin
contours correspond to galaxies without emission lines, and the thick
contours to emission-line galaxies. The bottom panel shows the $A_z$-$P2$
correlation.}
\label{etaEMLNOEML}
\end{figure}


\begin{figure}
\centering
\includegraphics[bb=80 57 532 759, width=\columnwidth]{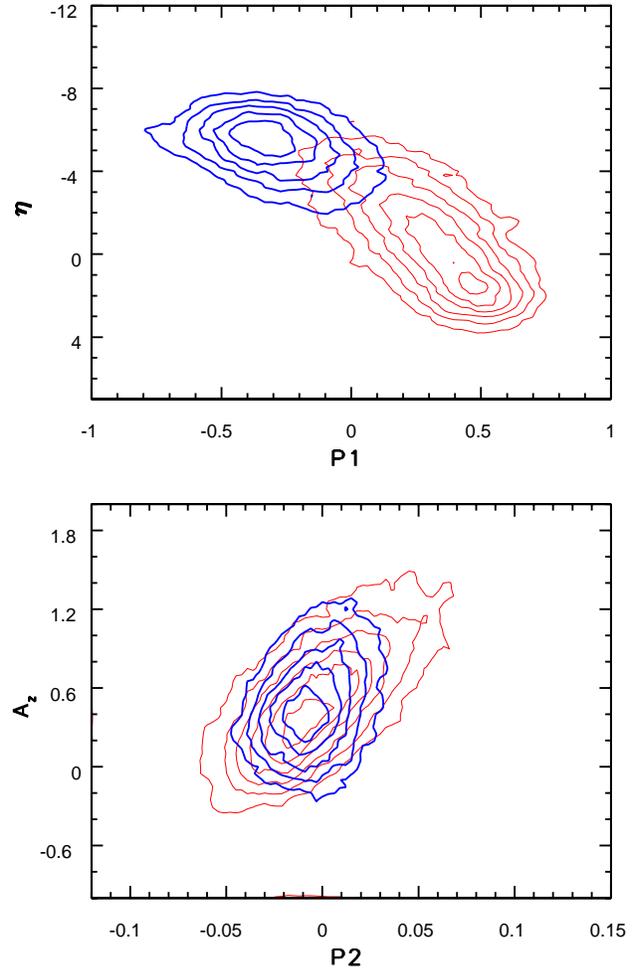}
\caption{Analogous to Fig.~\ref{etaEMLNOEML}, except that star-forming 
(thick contours) and AGN (thin contours) galaxies are shown.}
\label{etaSFAGN}
\end{figure}


\begin{figure}
\centering
\includegraphics[bb=80 57 532 759, width=\columnwidth]{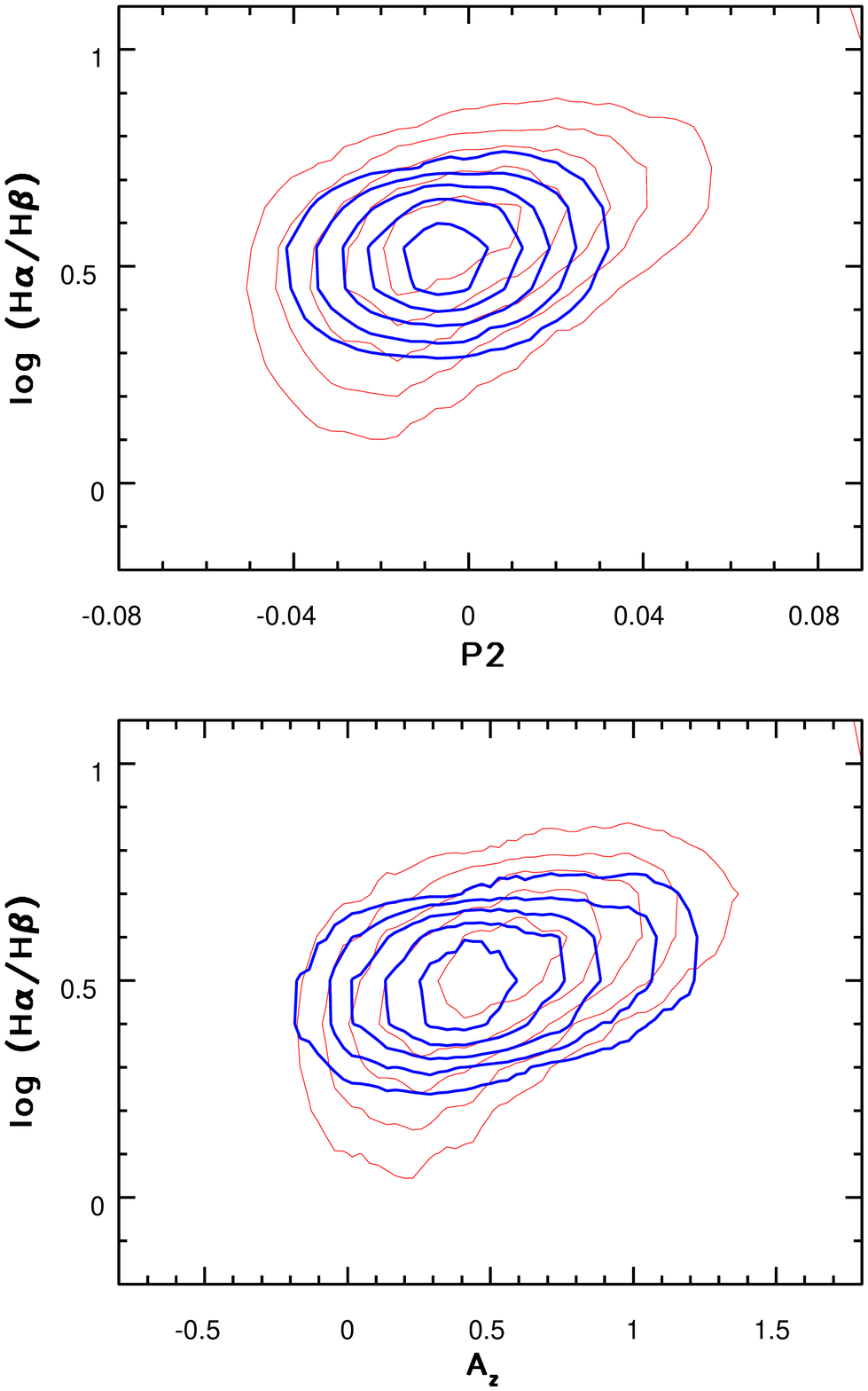}
\caption{The comparison of log($H_\alpha/H_\beta$) vs. $P2$ (top) and
log($H_\alpha/H_\beta$) vs. $A_z$ (bottom) correlations for star-forming 
(thick contours) and AGN (thin contours) galaxies}
\label{HH}
\end{figure}

Kauffmann et al.~(2003a) used the distribution of galaxies in the plane spanned
by the strength of the $H_\delta$ line and the 4000~\AA\ break ($D_n(4000)$) to obtain
model-dependent estimates of stellar masses and the dust content for SDSS galaxies. 
Given the position of a galaxy in the $H_\delta$--$D_n(4000)$ plane, the most 
probable mass-to-light ratio is drawn from a model library. With the measured
luminosity, this ratio then yields the stellar mass. The observed luminosity is
corrected for the dust extinction determined by comparing observed imaging
$g-r$ and $r-i$ colors to model-predicted colors (the latter do not include
the effects of dust reddening). The reddening correction needed to make
models agree with data is interpreted as an effective optical depth, $A_z$, 
due to the galaxy's interstellar dust. As discussed in Section~\ref{modelsSec},
Obri\'{c} et al. (2006) demonstrate that galaxies detected by IRAS have systematically 
higher values of $A_z$, supporting the notion that $A_z$ provides an estimate 
of the dust content. 

The analysis by Kauffmann et al. is based on the 3850--4160~\AA\ spectral range,
marked by the horizontal bar in the bottom panel in Fig.~\ref{fltsplot}, while 
the rest-frame Str\"omgren colors discussed here are determined using the
4000--5800~\AA\  range. That is, the two studies are based on practically independent 
spectral ranges (the $vz$ filter does have overlap with the $H_\delta$ line). 
Here we demonstrate that despite their different wavelength range,
and different techniques (color vs. spectral line analysis), the derived inferences
about galaxies are similar.

The top panel in Fig.~\ref{etaAll} shows the distribution of galaxies in the 
$H_\delta$-$D_n(4000)$ plane, with the dots colored according to the galaxy's position 
in the $P1$-$P2$ color-color diagram (see the top panel in Fig.~\ref{BPT}). As 
discussed by Kauffmann et al. (2003a), galaxies form a well-defined locus in 
this plane. The strong color gradient along the locus indicates a correlation 
with the $P1$ color. The bottom panel in Fig.~\ref{etaAll} shows the distribution 
of galaxies in the plane spanned by the derived parameters: the dust content and
the stellar mass. The color gradients indicate a good correlation between the positions
in the $A_z$-$M_\ast$ and $P1$-$P2$ planes. In particular, galaxies with small
masses (log($M_\ast$)$\la$10) are dominated by blue galaxies with $P1<0$,
while a correlation between $A_z$ and $P2$ is discernible (gradient from purple
to red as $A_z$ increases) for high-mass galaxies (log($M_\ast$)$\sim$11).
Since emission line properties, discussed in the previous Section, correlate
with $P1$ and $P2$ colors, it is expected that galaxies without emission 
lines, star-forming, and AGN galaxies display different distributions in these
two diagrams. This is demonstrated in Figs.~\ref{etaNOEML}, \ref{etaSF}, and
\ref{etaAGN}. For example, the detailed morphology of the galaxy distribution
shown in the bottom panel in Fig.~\ref{etaAll} can be understood as a superposition
of three dominant populations with distinctive positions, as seen in the 
bottom panels in Figs.~\ref{etaNOEML}, \ref{etaSF}, and \ref{etaAGN}.
It is noteworthy that the  $A_z$-$M_\ast$ correlation for 
star-forming galaxies shown in the bottom panel in Fig.~\ref{etaSF} is 
consistent with a constant dust-to-stellar mass ratio for all galaxies.

In order to quantify the correlations between the positions in the $H_\delta$-$D_n(4000)$ 
and $P1$-$P2$ planes, we define a parameter $\eta$ which measures the position 
along the locus in the $H_\delta$-$D_n(4000)$ plane
\eq{
     \eta =  0.0995\cdot (D_n(4000) - 1.75) - 0.995\cdot H\delta_A
}
Note that $\eta$=0 corresponds to $H\delta_A$=0. Fig.~\ref{etaEMLNOEML} compares 
the $\eta$-$P1$ and $A_z$-$P2$ distributions for galaxies with and without emission 
lines. As evident, there is good correspondence between the two pairs of parameters,
although they are measured using barely overlapping spectral ranges. In
particular, note the good correlation between $A_z$ and $P2$ for emission 
line galaxies. Analogous diagrams comparing star-forming and AGN galaxies
are shown in Fig.~\ref{etaSFAGN}. It appears that the $A_z$-$P2$ correlation
is stronger for AGN than for star-forming galaxies. This could be explained
as a consequence of the more homogeneous AGN population because 
star-forming galaxies span a broad range in stellar mass, and hence in metallicity 
and surface mass density. It also seems that the slope of the $A_z$-$P2$ correlation is 
different for the two types of galaxies, but the data are too noisy to derive a 
robust conclusion.

\subsection{   The Correlations of $A_z$ and $P2$ with the $H_\alpha/H_\beta$ Ratio}

\label{HaHb}

An often used estimator of the galaxy dust content is the $H_\alpha/H_\beta$ ratio.
Therefore, at least some degree of correlation should exist between this parameter
and the $A_z$ and $P2$ parameters. 
Fig.~\ref{HH} shows the dependence of log($H_\alpha/H_\beta$) on $P2$ and
$A_z$ for star-forming and AGN galaxies. Similarly to $A_z$ vs. $P2$ 
dependence shown in Fig.~\ref{etaSFAGN}, it appears that different weak 
correlations exist for these two types of galaxies. The difference in 
slopes, similar to the difference in slopes for the $A_z$ vs. $P2$ correlation, 
is possibly due to different dust geometries, or perhaps different
dust properties, between AGN and star-forming galaxies (note that this difference
also exists in the  $H_\alpha/H_\beta$ vs. $A_z$ diagram, that is, even
when $P2$ is not used). A detailed analysis and interpretation of these correlations 
is beyond the scope of this work. We conclude that the similarity of the top and 
bottom panels in Fig.~\ref{HH} suggests that $P2$ is a good, though somewhat noisier, 
proxy for $A_z$.

\subsection{         The Color -- Absolute Magnitude Distribution }

\label{MrSec}


\begin{figure}
\centering
\includegraphics[bb=150 67 452 759, width=\columnwidth]{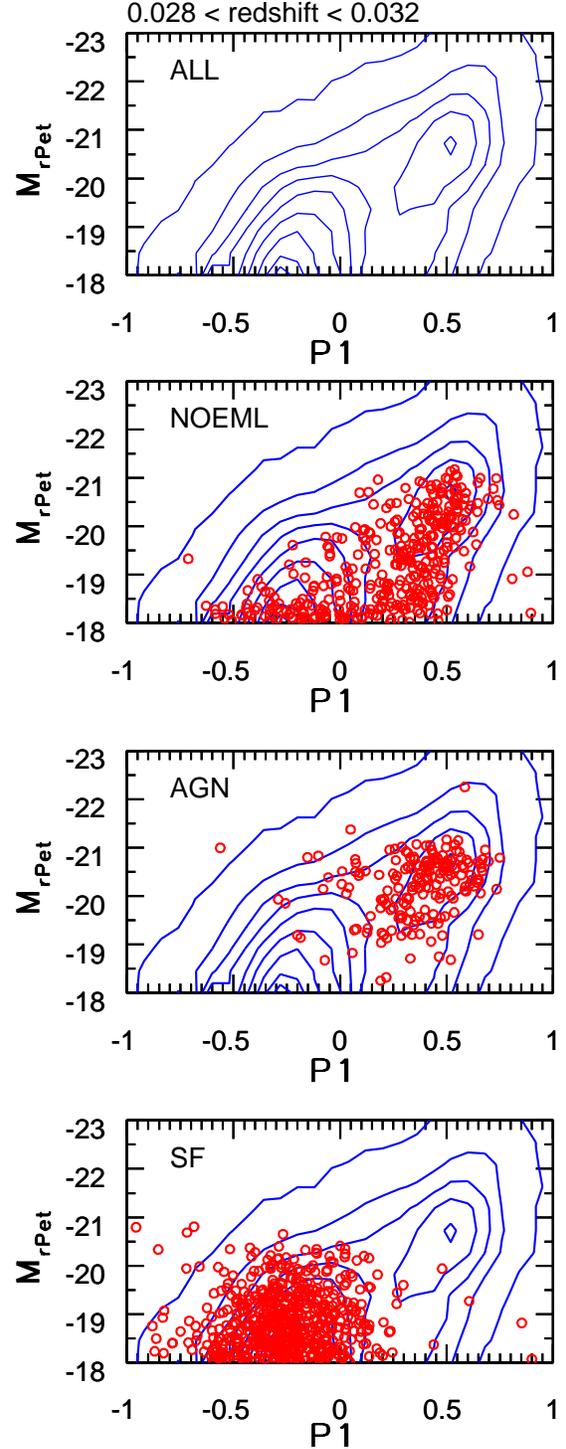}
\caption{The color-absolute magnitude diagram for SDSS main galaxies.
The top panel shows the absolute $r$ band magnitude as a function
of the P1 rest-frame Str\"omgren color for galaxies with 0.028$<$redshift$<$0.032.
The remaining panels compare the distribution of all galaxies (contours) to the
distributions of three subsamples (symbols) selected using emission lines: 
galaxies without emission lines in the upper middle panel, AGN galaxies 
in the lower middle panel, and star-forming galaxies in the bottom panel.
}
\label{MrP1}
\end{figure}

The color distribution of galaxies depends on their absolute magnitude:
the fraction of red galaxies increases with luminosity. Furthermore,
the median color of red galaxies becomes even redder with luminosity
(Baum 1959, Faber 1973). These long-known results are spectacularly
confirmed by the accurate and voluminous SDSS data (e.g. Blanton et al. 2003, 
Baldry et al. 2004, and references therein). Hence, the distribution of the 
$P1$ principal color introduced here is expected to be highly correlated with 
luminosity. 

The top panel in Fig.~\ref{MrP1} shows the distribution of 1,704
galaxies selected from a narrow redshift range in the absolute magnitude
vs. $P1$ color plane. We use the Petrosian $r$ band magnitudes without
K-correction and WMAP cosmology (Spergel et al. 2003) to compute $M_{rPet}$ 
(for a description of Petrosian magnitudes see Stoughton et al. 2002, and Strauss 
et al. 2002). The strong color-magnitude relation for red galaxies is evident, 
as well as the dramatic increase in the fraction of red galaxies for 
$M_{rPet} < -20$. Indeed, red galaxies with $P1\sim0.5$ show a well-defined
local maximum in their luminosity function (at $M_{rPet}\sim-20.7$). 

This figure suggests that $P1=0.25$ provides a good  separation between 
blue and red galaxies ($u-r=2.22$ separator from Strateva et al. 2001 
corresponds to median $P1=0.18$). The separation of less-luminous blue galaxies 
and more-luminous red galaxies is much more pronounced in this diagram based on 
the $P1$ color, than when using other colors such as K-corrected broad-band SDSS colors 
(see e.g. Fig. 7 in Blanton et al. 2003). This is presumably because $P1$ 
is a principal axis in a rest-frame color-color diagram constructed with 
narrow-band filters (the galaxy locus in the $r-i$ vs. $g-r$ rest-frame
color-color diagram, discussed in Section \ref{riSec}, is twice as broad as 
the galaxy locus in the $bz-yz$ vs. $vz-yz$ plane).

Since the $P1$ color is correlated with numerous other parameters, 
the subsamples separated using those parameters should have distinct
distributions in the color -- absolute magnitude diagram. As an example, 
we choose the separation into star-forming, AGN and galaxies without 
emission lines. The bottom three panels in Fig.~\ref{MrP1} compare the 
color--absolute magnitude distribution for each subsample to that for 
the full sample. As evident, AGN and galaxies without emission lines are 
predominantly red and luminous, while star-forming galaxies are blue and 
faint. The faint blue galaxies without emission lines, visible in the upper 
middle panel, are galaxies with weak emission lines that did not fulfill 
the condition of the $3\sigma$ detection for all four relevant emission lines.

\subsection{         The Color -- Mass Distribution }

\label{massSec}


\begin{figure}
\centering
\includegraphics[bb=150 67 452 759, width=\columnwidth]{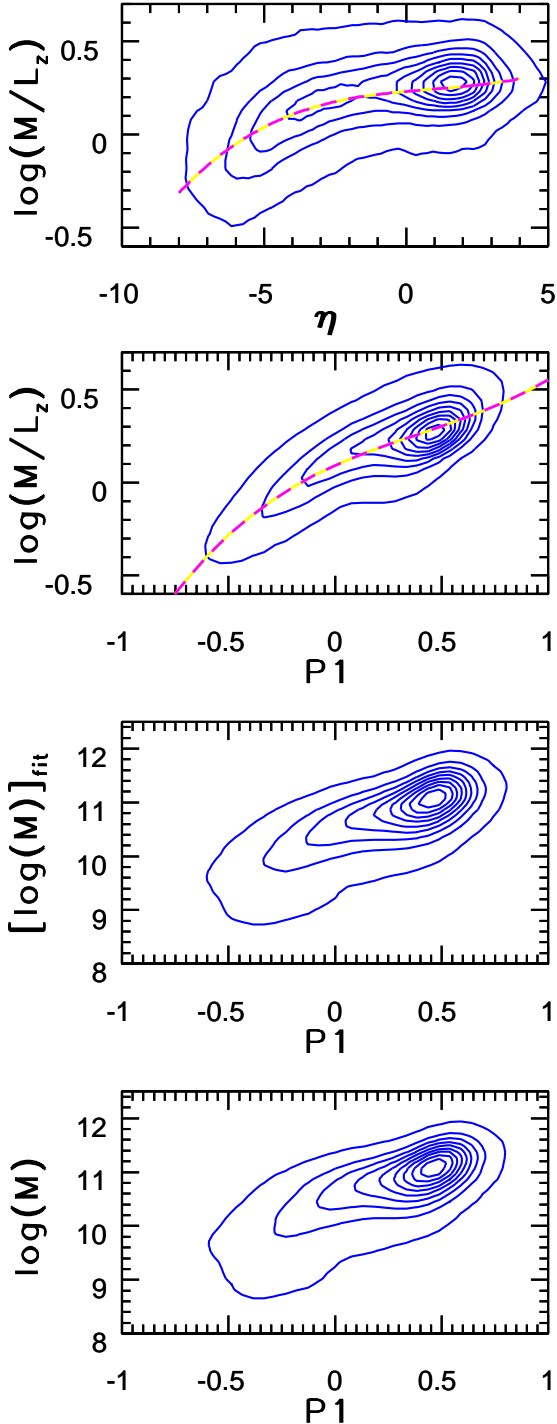}
\caption{The top panel shows the correlation between the mass-to-light ratio 
(in solar units) and
the position of a galaxy on the $H_\delta$ -- $D_n(4000)$ locus, where 
the model-dependent mass-to-light ratio is taken from Kauffmann et al. (2003). 
The dashed line is a best-fit third order polynomial. The upper middle panel 
shows the correlation between the mass-to-light ratio and the position of 
a galaxy on the locus in the rest-frame Str\"omgren color-color diagram. 
The dashed line is a best-fit third order polynomial. The lower middle panel 
shows the mass -- color diagram using the best-fit mass-to-light ratio determined 
from the $P1$ color, and the bottom panel shows analogous distribution using 
the mass-to-light ratio from Kauffmann et al. (2003). Note the similarity
of the two distributions.}
\label{massP1}
\end{figure}

Kauffmann et al. (2003a) estimated stellar masses for SDSS galaxies
by multiplying observed luminosities (corrected for extinction using $A_z$)
with the stellar mass-to-light
ratio inferred from the position of a galaxy in the $H_\delta$-$D_n(4000)$ 
plane. The correlation of the mass-to-light ratio and parameter $\eta$,
which measures the position of a galaxy along the $H_\delta$-$D_n(4000)$ 
locus, is shown in the top panel in Fig.~\ref{massP1}. The dashed line
is a best fit
\eq{
  \log{(M/L_z)}=0.000630\eta^3-0.00174\eta^2+0.01396\eta+0.231,
}
which ``predicts'' $\log{(M/L_z)}$ with an rms of 0.13 (the overall distribution
of $\log{(M/L_z)}$ has an rms of 0.17, i.e. a factor of 1.5). The upper middle
panel shows  the mass-to-light ratio as a function of the $P1$ color. 
The best fit, 
\eq{
   \log{(M/L_z)} = 0.273\,P1^3-0.328\,P1^2+0.516\,P1+0.091,
}
reproduces $\log{(M/L_z)}$ with an rms of 0.10, equal to the median error
in estimating $\log{(M/L_z)}$ using the Kauffmann et al. (2003a) method.
Hence, {\it it is possible to use $P1$ to convert the observed luminosities 
into stellar masses} with practically the same accuracy as the $H_\delta$-$D_n(4000)$
method employed by Kauffmann et al. (2003a). The bottom two panels compare 
the mass-color distributions when the mass is determined using $P1$, and 
when the mass is taken from Kauffmann et al. (2003a); their similarity
confirms this conclusion. We note that the overall 
appearance of the  mass-color distribution, in particular, the strong correlation 
of mass and color, essentially reflects the distribution of galaxies in the
color-luminosity diagram (because the mass-to-light ratio has a very
small dynamic range). 

Since $P1$ and luminosity, $L_z$, are correlated, as described in the 
previous Section, and $M/L_z$ is a function of $P1$, $M/L_z$ and $L_z$ should 
also be correlated to at least some extent. For example, Bernardi et al. (2003b) 
determined that for elliptical galaxies $M/L \propto L^{0.14}$. We find that a 
best power-law fit for {\it all} galaxies is $M/L \propto L^{0.4}$. It reproduces
$\log{(M/L_z)}$ with an rms of 0.20. Since this scatter is larger than
when determining $M/L$ from $P1$, we speculate that galaxy colors are
a better indicator of the mass-to-light ratio, and that its correlation 
with galaxy luminosity is probably a consequence of the color-magnitude 
relation.

\subsection{         The Color -- Velocity Dispersion Distribution }

\label{vdispSec}

\begin{figure}
\centering
\includegraphics[bb=150 327 452 759, width=\columnwidth]{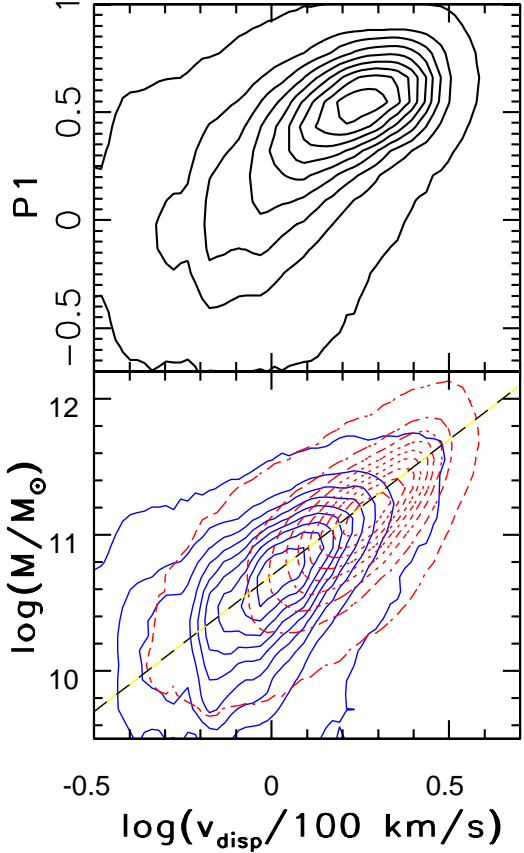}
\caption{The top panel shows the correlation between the P1 rest-frame Str\"omgren 
color and the velocity dispersion. The bottom panel shows the correlation between
the galaxy's stellar mass and the velocity dispersion, separately for galaxies
with P1$<$0.25 (solid lines) and P1$>$0.25 (dashed lines). The diagonal straight 
line, log($M/M_\odot$)=10.7+2\,log($v_{\rm disp}$/100~km/s), is added for
illustration}
\label{vdispP1}
\end{figure}

Stellar velocity dispersion is automatically measured for each SDSS galaxy
by fitting the observed spectrum with a linear combination of galaxy 
template spectra broadened by a Gaussian kernel (Schlegel et al., in prep.).
Detailed analysis of velocity dispersions for elliptical galaxies, as measured 
by SDSS, is discussed by Bernardi et al. (2003a) and Sheth et al. (2003).

We find a good correlation between the velocity dispersion and the $P1$ color,
shown in the top panel in Fig.~\ref{vdispP1}. The correlation is especially
strong for velocity dispersions larger than the instrumental resolution
of the SDSS spectra (60-70~km/s, log($v_{\rm disp}$/100~km/s)\about\,-0.2).
A similar correlation between the velocity dispersion and the $g-r$ color
for elliptical galaxies is discussed by Bernardi et al. (2003d).

The correlation between the velocity dispersion and the $P1$ color appears to
be a consequence of the good correlation between the stellar mass and 
the velocity dispersion, shown in the bottom panel in Fig.~\ref{vdispP1}. 
We find that the function
\eq{
 \log{(M/M_\odot)}=10.7+2\,\log\left({v_{\rm disp} \over 100\,{\rm km/s}}\right).
}
provides a good description for {\it both} blue and red galaxies. 
For a constant $M$, in the range $10.5 < \log{(M/M_\odot)} < 12$, the rms scatter
in the velocity dispersion is 40~km/s, independent of $M$. For a constant
velocity dispersion, the rms scatter in $\log{(M/M_\odot)}$ decreases from 
0.33 for $v_{\rm disp}=100$~km/s to 0.18 for $v_{\rm disp}=250$~km/s.  

The similar $\log{(M/M_\odot)}$ vs. $v_{\rm disp}$ behavior for blue and red 
galaxies can be considered as circumstantial evidence that the model-dependent 
mass-to-light ratios must be correct to some extent. The reason is that 
blue ($P1<0.25$) and red galaxies ($P1>0.25$) have different observed 
luminosities (median $z$ band luminosity for red galaxies is larger by a 
factor of 2.6 than for blue galaxies), and different velocity dispersions, and only 
the ``correct'' mass-to-light ratios (medians are 1.2 and 1.9 for blue and
red galaxies, respectively) will place the inferred stellar masses on 
the same $M-v_{\rm disp}$ relation. If the same mass-to-light ratio is 
used for all galaxies, the inferred mass for red galaxies would be smaller 
by a factor of 1.6 than for blue galaxies with the same velocity dispersion.
That is, the two distributions shown in the bottom panel in 
Fig.~\ref{vdispP1} would be vertically offset by \about0.2. The data constrain 
the offset between the two distributions to $\la$0.1.

\subsection{         The Rest-frame Str\"omgren Colors vs. Red Colors}

\label{riSec}


\begin{figure}
\centering
\includegraphics[bb=150 67 452 759, width=\columnwidth]{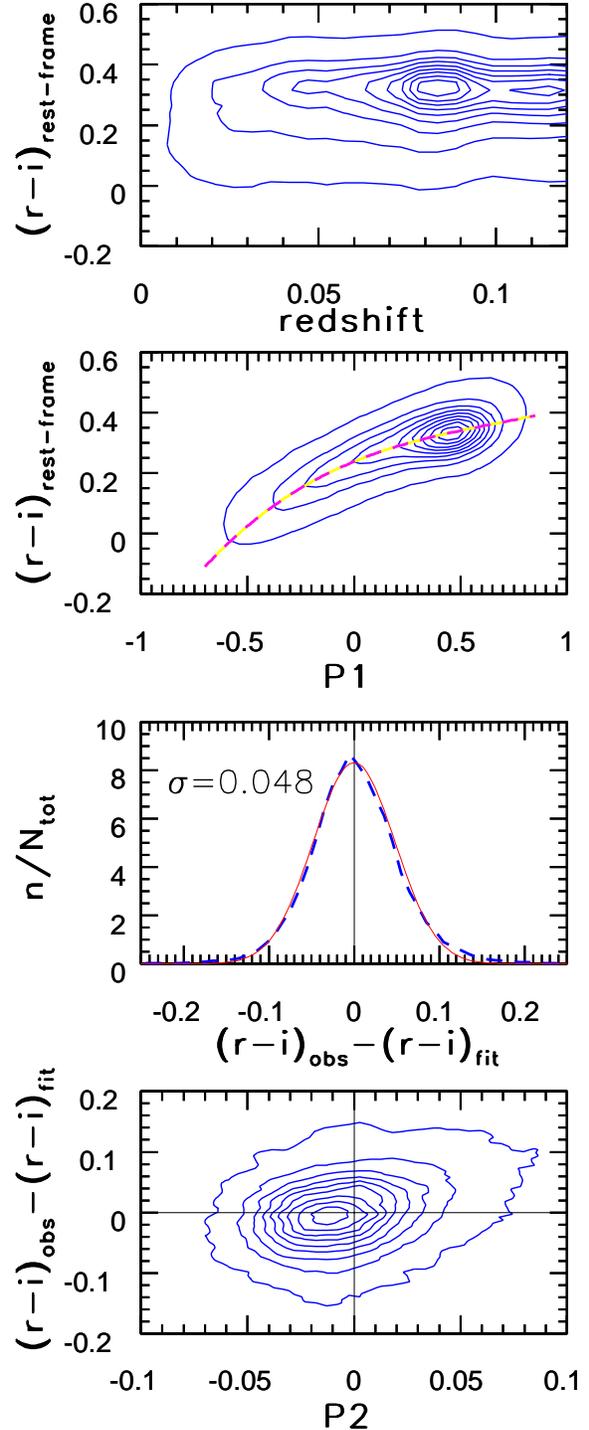}
\caption{The top panel shows the rest-frame $r-i$ color as a function of 
redshift. The upper middle panel shows the correlation between the rest-frame 
$r-i$ color and the P1 rest-frame Str\"omgren color. The dashed line is a 
best-fit third order polynomial (see text). The distribution of residuals between 
the fit and measured values is shown in the lower middle panel by the thick
dashed line. The thin solid line is a Gaussian with the width of 0.048~mag.
The bottom panel shows the residuals as a function of the P2 rest-frame 
Str\"omgren color.}
\label{P1ri}
\end{figure}

We have demonstrated in the preceding sections a good correlation
between rest-frame Str\"omgren colors and parameters determined
from the blue part of the spectrum ($H_\delta$ and $D_n(4000)$). In this
section we analyze the correlation between rest-frame Str\"omgren colors 
and parameters determined from the red part of spectrum. As shown in 
Section~\ref{modelsSec}, the age-metallicity degeneracy cannot be broken
using rest-frame Str\"omgren colors employed here. It is possible
that additional information contained at the redder wavelengths could 
provide a method to separate the age and metallicity effects.
If the properties of the red part of the spectrum can be fully predicted 
using rest-frame Str\"omgren colors, then such a possibility can be
ruled out. Furthermore,
if the blue and red parts of the spectrum are dominated by uncorrelated
stellar populations, it should {\it not} be possible to accurately predict
$r-i$ color using blue Str\"omgren colors.

To characterize the red part of the spectrum, we synthesize the {\it rest-frame}
$r-i$ color, using the same method as for synthesizing rest-frame Str\"omgren 
colors. The top panel in Fig.~\ref{P1ri} shows the rest-frame $r-i$ color
as a function of redshift. Reassuringly, there is no strong dependence on 
redshift -- the upper limit on the change of the median color with redshift 
is \about0.02~mag (in the 0.02--0.12 redshift range). The correlation
between the rest-frame $r-i$ color and the $P1$ color is shown in the 
upper middle panel. The dashed line is a best-fit 
\begin{equation}
  r-i = 0.109\,P1^3 - 0.221\, P1^2 + 0.288\,P1 + 0.238,
\end{equation}
which predicts the rest-frame $r-i$ color with an rms of only \about0.05~mag. The 
distribution of residuals is shown in the lower middle panel, and is
well described by a Gaussian distribution.

Since $P2$ is a measure of dust reddening in a galaxy, some correlation 
between the $r-i$ residuals and $P2$ could be expected, although
the errors in the determination of $P1$, $P2$ and $r-i$, which are comparable 
to the magnitude of $r-i$ residuals, may easily mask it. The dependence of 
residuals on the $P2$ color, shown in the bottom panel, gives a hint of 
such a correlation. For example, the median $r-i$ residual is 0.03~mag
for $P2$=0.05, while it is -0.02~mag for $P2$=-0.05.

This one-dimensional parameterization of galaxy spectral energy distributions 
(SEDs) extends to even longer wavelengths. Obri\'{c} et al. (2004, 2006) find that 
SDSS imaging magnitudes in the $u$ and $r$ bands, together with redshift 
information, can be used to predict 2MASS $K$ band (2.2 \mic) magnitudes 
with an rms scatter of only \about0.2~mag, with a significant fraction of this 
scatter contributed by measurement errors (they estimate that astrophysical 
scatter is  much smaller than 0.1~mag). In summary, {\it the UV to IR SEDs of galaxies can be
described as a single-parameter family with an accuracy of 0.1~mag, or 
better; in addition, the more detailed spectral diagnostic parameters, such 
as line strengths, appear well correlated with the overall SED.}

\section{ The Correlation Between the Rest-frame Str\"omgren Colors and Spectral Eigencoefficients }

In the previous Section we discussed correlations between the rest-frame 
Str\"omgren colors and various other spectral parameters. Here we study 
the correspondence between a classification scheme based on spectral
eigencoefficients proposed by Yip et al. (2004) and the principal colors 
P1 and P2.

Yip et al. used the Karhunen-Lo\'{e}ve transform to classify 170,000 
SDSS galaxy spectra. These spectra are classified in a plane spanned by
the mixing angles between the first three eigencoefficients, $\Phi_{KL}$
and $\Theta_{KL}$. This classification discriminates between early-, 
late-type and extreme emission line galaxies (see their Section 5 for 
details). Given correlations between various spectral parameters discussed
here, it is expected that the distribution of galaxies in the 
$\Phi_{KL}$--$\Theta_{KL}$ plane is, at least to some extent, correlated with 
the principal colors  P1 and P2.

\begin{figure}
\centering
\includegraphics[bb= 60 50 550 750, width=\columnwidth]{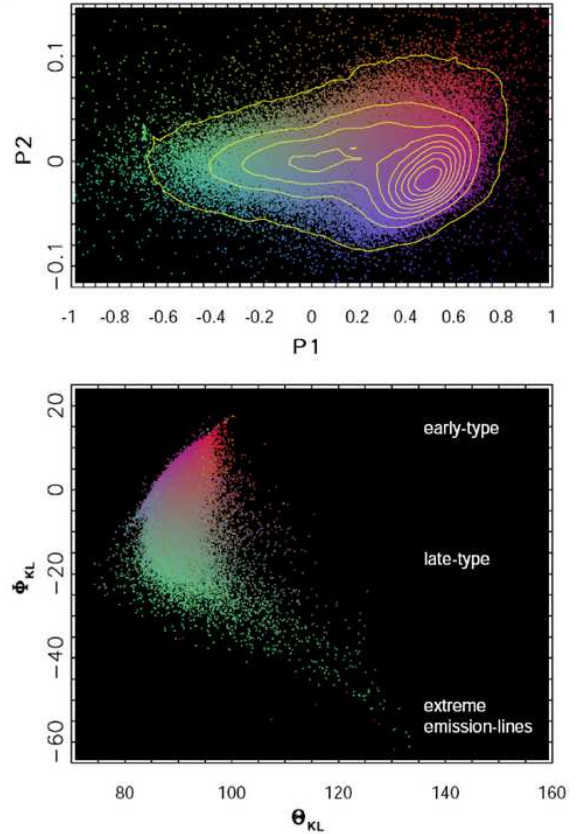}
\caption{The bottom panel shows a spectral classification plane for galaxies 
proposed by Yip et al. (2004). Each galaxy is represented by a dot color-coded 
according to its position in the P1-P2 plane, shown in the top panel.
The sequence from early-type to extreme emission line galaxies adopted from 
Yip et al. (2004) is shown on the right side in the bottom panel. Note
the good correspondence between P1 and P2 colors and the Yip et al. 
classification.}
\label{YipP1P2}
\end{figure}

Fig.~\ref{YipP1P2} shows this distribution, color coded according to the 
position of each galaxy in the P1-P2 plane. The correlation between the P1 
color and the proposed classification is evident. As discussed in detail by 
Yip et al., the mean spectra for galaxies selected from small 
($\Phi_{KL}$, $\Theta_{KL}$) regions correspond to different spectral 
types. This suggests that the P1 color is a robust measure of 
galaxy type, and confirms that most of the variance in galaxy SEDs is captured
by a small number of eigencomponents, as found by Yip et al.

The correlation with P2 is not as strong as for P1, but appears discernible 
for early- to late-type galaxies. There is a slight gradient from purple ($P2<0$) 
to red-orange ($P2>0$) in the $80<\Theta_{KL}<100$ range.


\section{ The Rest-Frame UV Colors of $\z\gtrsim 0.18$ Galaxies}
\label{sec:uz}

\begin{figure}
\centering
\includegraphics[bb= 80 200 530 520, width=\columnwidth]{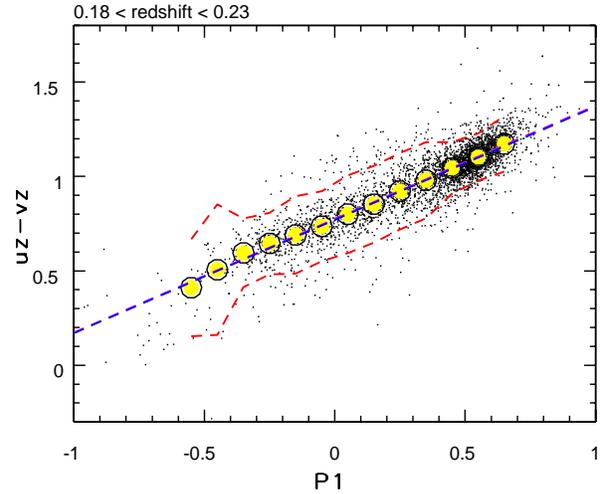}
\caption{The \UV~ color vs. P1 color-color diagram for galaxies (dots) with 
synthesized \U~magnitudes (0.18 $<$z$<$0.23). Large circles show medians
for P1 bins. The middle (straight) dashed line is the best-fit to these
medians, and the other two dashed lines mark the 2$\sigma$ envelope around
the medians. The best-fit relation can be used to ``predict'' the \UV~ color 
from P1 with a residual scatter of only 0.08~mag. 
}
\label{uzP1}
\end{figure}

Our analysis of the rest-frame Str\"omgren colors in the preceeding sections 
did not utilize the $uz$ magnitudes. Since the SDSS spectra cover only wavelengths
longer than $3800~\AA$, the \U~magnitudes can be synthesized only for 
galaxies at redshifts greater than $z\sim0.18$ (hereafter ``high-z'' sample). 
The ``high-z'' sample includes only  $\sim 9\%$ of the 99,088 galaxies from 
the ``main'' galaxy sample, and is strongly biased towards red galaxies 
because of the severe luminosity cut (see Fig.~\ref{MrP1}). 
Nevertheless, it includes a sufficient number of blue galaxies to demonstrate
a correlation between the \UV~color and the P1 color, shown in Fig.~\ref{uzP1}. 

The best-fit relation shown in Fig.~\ref{uzP1}, 
\begin{equation}
  uz-vz = 0.60\,P1 + 0.77,
\end{equation}
can be used to ``predict'' the \UV~color with a residual scatter of only 0.08~mag!
It is noteworthy that the \UV~residuals are not correlated with the P2 color.
This small scatter means that either the dependence of the \UV~color on
metallicity is not as strong as assumed a priori, or that age and metallicity
for galaxies are combined in such a degenerate way that the distribution 
of galaxies in the \UV~vs. P1 plane is essentially one-dimensional (within
this 0.08~mag scatter), similarly to their distribution in the P1 vs. P2 plane.

\begin{figure}
\centering
\includegraphics[bb= 20 200 590 600, width=\columnwidth]{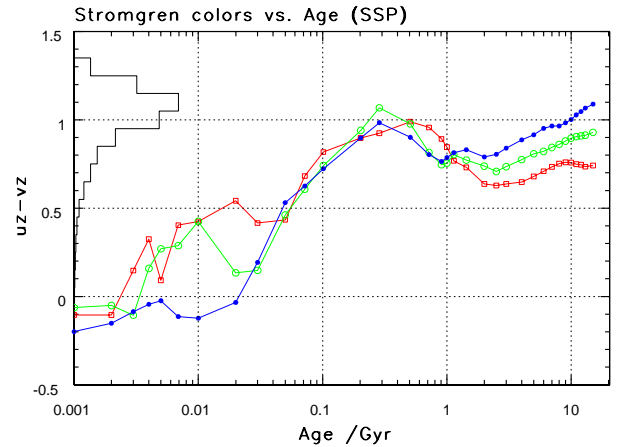}
\caption{The Bruzual~\& Charlot model predictions for the dependence of the
  \UV~color on age and metallicity ([Z]=0.004, squares; 0.02, circles;
  0.05, dots) of single stellar populations. Note that for small ages ($\la$30 Myr)
metallicity can have a strong impact on the \UV~color, while for intermediate
ages (30 Myr $\la$age $\la$ 2~Gyr) it plays only a minor role. For old 
stellar populations ($\ga$2~Gyr), the \UV~color becomes again increasingly 
sensitive to metallicity. However, not even for the oldest galaxies the model
\UV~color is sufficiently red to explain the observations.} 
\label{uzAge}
\end{figure}

In order to distinguish between these two possibilities, we resort again to 
the Bruzual~\& Charlot population synthesis models (see Section~\ref{modelsSec}). 
Fig.~\ref{uzAge} shows model predictions for the dependence of the \UV~color 
on age and metallicity. At least for the smallest ($\la$30 Myr) and 
largest ($\ga$2~Gyr) ages, the \UV~color appears sensitive to metallicity.
However, this ``high-z'' sample doesn't include extremely blue galaxies
with implied ages below 30 Myr, or P1$<$-0.5 (see Fig.~\ref{BCmodels}), 
so model predictions are not at odds with the correlation displayed in 
Fig.~\ref{uzP1}. The situation is less clear for old red ($\ga$2~Gyr, or P1$\ga$0.2) 
galaxies. Model tracks for 10~Gyr old galaxies (P1$>$0.4) predict a
$>$0.2-0.3~mag wide range of \UV~colors, while the observed distribution
has the root-mean-square scatter of 0.08~mag. 

\begin{figure}
\centering
\includegraphics[bb= 80 200 530 520, width=\columnwidth]{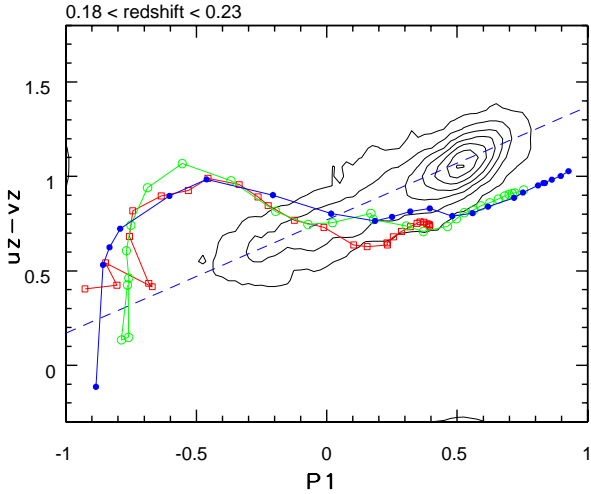}
\caption{The \UV~ color vs. P1 color-color diagram. The observed galaxy 
distribution is shown by linearly spaced contours (the same distribution is 
shown by dots in Fig.~\ref{uzP1}). The Bruzual~\& Charlot model predictions 
are shown by lines and symbols, analogously to Fig.~\ref{uzAge}. The position
along each model track is controlled by the age of a single stellar population,
while different tracks correspond to different metallicities. Note that
the effects of age and metallicity are largely degenerate such that different
model tracks do not deviate from each other by more than $\sim$0.1~mag. 
Note also the discrepancy between the model tracks and the observed distribution.
}
\label{uzP1model}
\end{figure}

The resolution of this discrepancy is the degenerate dependence of
the P1 and \UV~colors on age and metallicity. For galaxies older than 
$\sim$2~Gyr, {\it both} colors become redder as age and metallicity
increase, by about the same amount. As shown in Fig.~\ref{uzP1model},
for P1$>$0.4 the model tracks converge and become a single track 
irrespective of metallicity. Taken at face value, this convergence
suggests that the \UV~color scatter of elliptical galaxies (at a given P1) 
is due to age spread, and not due to a finite width of their metallicity 
distribution.

This interpretation has to be taken with caution because models never produce 
sufficiently red \UV~colors to explain the observations of the reddest 
galaxies. While one is tempted to invoke the dust (which is not incorporated 
in these models) as an explanation of this discrepancy, these galaxies are 
the reddest and most luminous galaxies which are not supposed to have 
significant amounts of dust. Furthermore, the models show significantly
different behavior for blue galaxies than displayed by the data distribution. 
It is quite plausible that model spectra in the rest-frame UV 
wavelength range require significant improvements. For example, 
Fig.~\ref{uzP1model} indicates that a model spectrum fit to $\lambda>4000$~\AA\
spectral range for a blue galaxy (say, P1=$-0.5$), will have the flux
in the 3200--4000~\AA\ wavelength range underestimated by $\sim$50\%
(0.5~mag). Note that this discrepancy cannot be easily attributed 
to dust effects (which are not accounted for in model spectra) because
model spectra are redder, and not bluer, than the observed spectra.


\section{ Discussion and Conclusions}

We analyzed the synthesized rest-frame Str\"omgren colors for \about100,000 
SDSS galaxies. This narrow-band photometric system is well suited for studying
galaxy colors in the 4000--5800~\AA\ wavelength range. Galaxies form a narrow
locus in the rest-frame Str\"omgren $bz-yz$ vs. $vz-yz$ color-color diagram, with a width of
only 0.03~mag. This small width, in addition to demonstrating that the errors 
in synthesized colors are small, shows that the slope of galaxy spectra in the 
3200--5800~\AA\ wavelength range is practically a one-parameter family. This 
conclusion is independent of the details of the employed photometric system.

Using the Bruzual~\& Charlot population synthesis models, we find that the 
dependence on age and metallicity is {\it fully absorbed} in the first principal 
color axis, $P1$. This color is essentially a measure of the Hubble type
(in sense of early vs. late type galaxies, rather than detailed morphological
classification). It is not possible to break the age-metallicity degeneracy with 
any of the observables discussed here. This may not be possible even when using 
standard methods such as Lick indices, at least in the case of elliptical galaxies 
(Eisenstein et al. 2003). As demonstrated by Eisenstein et al., the strengths 
of numerous metal lines, such as Mg b, can be accurately predicted from the 
strength of the $H_\delta$ line (see their Fig. 5); hence, no additional information 
can be extracted from the measurements of those lines. On the other hand, 
Tremonti et al. (2004) report a very tight mass-metallicity relation (0.1 dex
scatter in metallicity at a given mass) for \about50,000 SDSS star-forming 
galaxies. Thus, it may be that age and metallicity are strongly correlated, 
too, which then would provide a trivial solution to the problem of breaking the
age-metallicity degeneracy. Indeed, such a relation has been claimed for disk
stars in our Galaxy (Twarog 1980; however, see also Feltzing, Holmberg \& 
Hurley 2001).

The second principal color axis, $P2$, is independent of stellar
age and metallicity, at least according to the Bruzual~\& Charlot models.
We stress that the definition of this color is purely {\it observational}, 
and the models explain both its value and narrow distribution width. The $P2$ distribution
width (0.03~mag) measured using SDSS data is smaller than typical observational 
errors in rest-frame colors using other data available to date ($\ga$0.05~mag). 
The $P2$ color appears most sensitive to the galaxy dust content, as demonstrated 
by the analysis of model-dependent estimates of the dust content by Kauffmann et al. 
(2003a), and also by the redder $P2$ colors for galaxies detected by the IRAS survey. 
Of course, $P2$ could be {\it indirectly} related to metallicity, if the dust content 
depends on metallicity (that is, the observed $P2$ color could be correlated with
metallicity as a result of radiative transfer through interstellar dust, rather 
than due to the dependence of the source function on metallicity). 

The principal colors are well correlated with numerous other spectral parameters
determined from bluer and redder spectral ranges, as well as with the position 
of a galaxy in the BPT diagram constructed with emission-line strength ratios.
These good correlations suggest that not only are the overall UV-IR SEDs a nearly  
one-parameter family, but that the same parameter also controls the properties
of emission lines. Most notably, there is a good correspondence between the stellar 
mass-to-light ratio estimates by Kauffmann et al. (2003a) and the principal color $P1$.

The low dimensionality of galaxy spectra implied by the numerous correlations among 
the continuum and spectral parameters shown here is in good agreement with the 
conclusions by Yip et al. (2004), who applied the principal component analysis to 
SDSS galaxy spectra. The same conclusion was reached in a detailed study of elliptical 
galaxies by Eisenstein et al. (2003). These conclusions are valid for the overwhelming 
majority of all galaxies; however, we emphasize that the analysis presented here
is based on low-order statistical moments (such as root-mean-square scatter) and
cannot rule out a possibility that as many as 5-10\% of galaxies behave differently.

Our results support the suggestion by Rakos and collaborators (Fiala, Rakos \& Stockton 1986) 
that the rest-frame Str\"omgren photometry provides an efficient tool for studying faint cluster 
galaxies and low surface brightness objects, without the need for time-consuming spectroscopy.
For example, the stellar mass-to-light ratio can be determined from $P1$ with
an rms scatter of only 0.10. Since the brightest cluster galaxies can be selected 
from SDSS data to redshifts of \about 0.8 (e.g. Ivezi\'{c} et al. 2002), this method 
can be used to quantify the rise in star formation activity with cosmic epoch, 
and the Butcher-Oemler effect, with an unprecedented accuracy. We are currently 
undertaking such an observational program.

\section*{Acknowledgments}

V.S. and \v{Z}.I. are thankful to the Princeton University for generous financial
support. We thank Ching-Wa Yip for providing us galaxy classification results, 
and for useful discussions.

Funding for the SDSS and SDSS-II has been provided by the Alfred
P. Sloan Foundation, the Participating Institutions, the National
Science Foundation, the U.S. Department of Energy, the National
Aeronautics and Space Administration, the Japanese Monbukagakusho, the
Max Planck Society, and the Higher Education Funding Council for
England. The SDSS Web Site is http://www.sdss.org/.
 
The SDSS is managed by the Astrophysical Research Consortium for the
Participating Institutions. The Participating Institutions are the
American Museum of Natural History, Astrophysical Institute Potsdam,
University of Basel, Cambridge University, Case Western Reserve
University, University of Chicago, Drexel University, Fermilab, the
Institute for Advanced Study, the Japan Participation Group, Johns
Hopkins University, the Joint Institute for Nuclear Astrophysics, the
Kavli Institute for Particle Astrophysics and Cosmology, the Korean
Scientist Group, the Chinese Academy of Sciences (LAMOST), Los Alamos
National Laboratory, the Max-Planck-Institute for Astronomy (MPA), the
Max-Planck-Institute for Astrophysics (MPIA), New Mexico State
University, Ohio State University, University of Pittsburgh, University
of Portsmouth, Princeton University, the United States Naval
Observatory, and the University of Washington.

\end{document}